# Theoretical study of nitrogen-vacancy spin coherence enhanced by vacancy aggregation in diamond using cluster-correlation expansion

Chikara Shinei[1,2,*]

[1] Department of Applied Physics, Institute of Pure and Applied Sciences, University of Tsukuba, Tsukuba, Ibaraki 305-8573, Japan

[2] Japanese-French Laboratory for Semiconductor Physics and Technology J-FAST, CNRS, University Grenoble Alpes, Grenoble INP, University of Tsukuba, Tsukuba, Japan

[*] e-mail: shinei.chikara.fb@u.tsukuba.ac.jp

---

## Abstract

Annealing a semiconductor makes its vacancies mobile; they aggregate into multi-vacancy complexes that often carry spin. Such centres are a magnetic-noise source for any spin qubit among them, in silicon and silicon carbide as well as in diamond, and they are accordingly blamed for the decoherence of negatively charged nitrogen-vacancy ($NV^-$) centres in irradiated and implanted diamond, with two coherence records credited to removing them. Cluster-correlation-expansion simulations in which the spins outside each cluster keep exchanging spin projections in pairs reproduce the measured Hahn-echo coherence time of $NV^-$ centres in a substitutional-nitrogen bath to within 12 %, inside the statistical uncertainty. Driven by published electron-paramagnetic-resonance parameters, the same calculation inverts the attribution. At a fixed paramagnetic spin density a multi-vacancy bath gives a Hahn-echo coherence time 3.0–6.1 times longer than a bath of isolated negative vacancies at 1 ppm, and 2.2–6.5 times longer from 0.01 to 10 ppm. A bath dephases the qubit because its spins exchange spin projections with one another, and two can exchange only if their transition frequencies match. A fine-structure splitting shifts those frequencies. Were every defect on the same crystallographic site, all would shift alike and still match, and the splitting would gain little, a factor 1.2. Real defects occupy symmetry-equivalent sites pointing in different directions, so neighbours land at different frequencies and stop exchanging: a factor 2.3. What governs the coherence time is therefore the fraction of bath pairs sharing a transition frequency, not the size of the zero-field splitting. Vacancy aggregation extends $NV^-$ spin coherence rather than shortening it.

---

## Introduction

A semiconductor that has been irradiated, implanted, or grown quickly contains vacancies, and annealing makes them mobile. They diffuse, meet and aggregate, and the multi-vacancy complexes left behind are often more stable than the isolated vacancy. Silicon is the best-studied case. Electron paramagnetic resonance identified the divacancy, trivacancy, tetravacancy and pentavacancy in irradiated crystals, and the same body of work traced the attractive force that binds vacancies into chains along ⟨110⟩ — the motif that recurs in the diamond centres studied below.[1] Tight-binding and first-principles calculations then find a series of especially stable sizes — $V_6$, $V_{10}$, $V_{14}$ and larger — set by how the dangling bonds at the vacant sites pair off,[2] and the decavacancy $V_{10}$ is predicted to hold high-spin states in several charge states, one of the few semiconductor defects for which that is so.[3],[4] Silicon carbide behaves the same way. Density-functional calculations map the stable multivacancies of 4H-SiC,[5] the neutral divacancy has a spin-triplet ground state that electron paramagnetic resonance detects after annealing near 850 °C,[6] and larger clusters such as the trivacancy V_C–V_Si–V_C appear as the divacancy anneals out.[7] In the two materials that carry most of semiconductor technology, as in diamond, aggregation therefore tends to leave centres with electron spin $S \geq 1$.

Those centres are a magnetic-noise source for any spin qubit sitting among them, so the coherence of a quantum-sensing defect is tied to what the surrounding vacancy population has aggregated into. This paper asks what that costs, or buys, in the case where the coherence is longest and best characterised: the negatively charged nitrogen-vacancy ($NV^-$) centre in diamond at room temperature.

$NV^-$ centres, in ensembles and one at a time, are the working element of a large class of solid-state magnetometers. The Hahn-echo coherence time $T_2$ sets their ac sensitivity, and it is the quantity studied here. $NV^-$ ensembles are normally made in nitrogen-rich diamond, where an electron spin bath of neutral substitutional nitrogen (P1, N_s⁰) limits it, and it scales as the inverse of the paramagnetic nitrogen concentration.[8],[9] The mechanism is well understood. The $NV^-$ dephases in the dipolar field of the P1 spins, and that field fluctuates because pairs of P1 centres exchange spin projections (flip-flop). A flip-flop conserves energy only if both partners share the same electron-paramagnetic-resonance (EPR) transition frequency. Anything that spreads the bath transition frequencies therefore slows the bath down. In a P1 bath the on-site $^{14}N$ hyperfine interaction provides that spread, acting along a Jahn–Teller axis that points at random along one of the four ⟨111⟩ directions. The cluster-correlation expansion (CCE) now reproduces this picture quantitatively. Park et al. computed the Hahn-echo $T_2$ of $NV^-$ ensembles in a P1 bath from 1 to 100 ppm and obtained a log–log slope of −1.06, consistent with the −1.07 that describes the measured dependence, and with no adjustable parameter.[10] The same work showed that the orientational disorder is part of the mechanism: a bath whose Jahn–Teller axes all lie along [111], or one with the hyperfine interaction switched off, decoheres the $NV^-$ markedly faster than the real, orientationally disordered bath.[10]

Irradiation or ion implantation followed by annealing leaves a very different defect inventory, and that inventory is what $NV^-$ centres made by implantation actually sit in. Ensemble EPR maps it. Above ~800 °C the vacancies are mobile and the divacancy has already annealed out. What remains is dominated by ⟨110⟩ multi-vacancy chains V_n⁰ ($n \geq 3$) — R5 ($V_3$), O1 ($V_4$), R6 ($V_5$), R10 ($V_6$), R11 ($V_7$), KUL11 ($V_8$) — with the shorter chains most abundant. Above ~1100 °C, closed multi-vacancy clusters such as R7, R8 and R12 take over.[11–13] Every one of these centres has $S = 1$ and a resolved fine-structure tensor, because its two unpaired electrons sit in dangling bonds at the two ends of the cluster. An isotropic broad signal accompanies them. It has no fine structure at all, and more than one species contributes to it, including amorphous-carbon damage.[11]

The prevailing reading blames the multi-vacancy complexes. Yamamoto et al. anneal at 1000 °C, obtain a long $T_2$, and attribute it to "a reduction of paramagnetic vacancy clusters near $NV^-$"; in samples annealed at 800 °C they identify those clusters as "the main source of decoherence of $NV^-$".[11] Herbschleb et al. report the longest room-temperature $T_2$ measured in any solid-state spin, 2.4 ms, in phosphorus-doped n-type diamond. They attribute it to "charging of the vacancies, which suppresses the formation of paramagnetic vacancy complexes during growth", and note that chemical-vapour-deposition growth generates many vacancies, which "cause generation of thermally stable impurity-vacancy and multi-vacancy complexes"; Coulomb repulsion between charged vacancies can suppress that.[14] A related result obtained with a sacrificial boron-doped layer is read the same way.[14] In each case the multi-vacancy complex is the species one wants to be rid of.

No one has tested that reading against a calculation of what a multi-vacancy bath actually does to an $NV^-$ spin. Park et al. list $V^-$, $[V–V]^0$ and related complexes among the parasitic spins that chemical-vapour-deposition growth and ion implantation generate, and note that their contribution to $NV^-$ decoherence is largely unknown.[10] This paper asks exactly that. If $NV^-$ centres sit in a multi-vacancy-limited bath, is their coherence worse than in a bath of isolated vacancies at the same paramagnetic density?

Symmetry suggests the answer goes the other way. The isolated negative vacancy $V^-$ (the S1 EPR centre) has $S = 3/2$ but keeps full T_d symmetry, and Isoya et al. noted that fine-structure splittings vanish for $S = 3/2$ in T_d.[15] Every $V^-$ in the crystal therefore shares one EPR transition frequency. Every pair is resonant, every pair can flip-flop, and spectral diffusion runs as fast as the dipolar coupling allows. A multi-vacancy complex is the opposite case. Its symmetry is lower, its fine-structure tensor has $D$ of tens to hundreds of MHz,[12],[13],[16] and — crucially — its principal axis lies along one of several symmetry-equivalent crystallographic directions. Two neighbouring multi-vacancy defects usually sit on differently oriented sites, and that detunes them from one another by far more than their dipolar coupling. Figure 1 sets out the model these two cases suggest; the rest of the paper tests it.

A second reason is purely combinatorial. Aggregation does not merely rearrange the spins; it removes them. $n$ isolated negative vacancies carry $n$ centres of $S = 3/2$, whereas the $n$-vacancy chain they form carries a single centre of $S = 1$, its two unpaired electrons sitting in the dangling bonds at either end. At fixed total vacancy content, aggregation therefore cuts the paramagnetic centre count by a factor $n$ and makes each surviving centre less damaging. Both effects push the same way. Neither appears in the picture that treats multi-vacancy complexes simply as a decoherence source to eliminate.

Defect-resolved spin-bath simulation now makes the question answerable. Park et al. have combined density-functional spin-Hamiltonian parameters with CCE to treat heterogeneous baths holding several paramagnetic species at once, and found that coherence depends on defect identity and bath composition rather than on total defect density alone.[17] Their vacancy-related species are the isolated vacancy and vacancy–hydrogen complexes; the multi-vacancy chains and clusters that dominate the annealed inventory have not been treated this way. This paper treats them, with the CCE.[18]

Conventional CCE freezes the bath spins outside each cluster, and for the P1 bath it overestimates the measured $T_2$. A dynamical-mean-field extension, in which those outer spins keep flip-flopping in pairs, removes most of that excess.[19] That calculation is used here throughout. Its only input is published EPR parameters, and it has no adjustable parameter. Four steps follow. First, the calculation is checked against the P1 bath, the one real spin bath for which measurements exist. Second, $T_2$ is predicted against concentration for every multi-vacancy centre considered. Third, a controlled decomposition isolates which property of the multi-vacancy bath is responsible. Fourth, the calculation is applied to a defect composition measured by EPR after annealing. The calculation reproduces the measured P1 Hahn-echo time within its statistical uncertainty. At a fixed paramagnetic spin density, every multi-vacancy bath gives a Hahn-echo coherence time 3.0–6.1 times longer than a bath of isolated negative vacancies at 1 ppm. The decomposition attributes most of that gain to the symmetry-equivalent site orientations; a fine-structure splitting with a single orientation gains only a factor 1.2. In the measured composition neither the multi-vacancy chains nor the accompanying isotropic species is the damaging part of the bath; the coherence follows the total spin count and the variety of the chain population. Vacancy aggregation therefore lengthens $NV^-$ spin coherence rather than shortening it.

---

## Results

### Benchmark on the P1 bath

Before the calculation is used to predict anything, it has to reproduce a bath that has been measured. The P1 bath is that case. Bauch et al. determined the Hahn-echo time across 25 nitrogen-doped diamonds and found $T_2$ = 160 ± 12 µs·ppm, as a slope against the P1 concentration.[8] The same field and protocol used for every multi-vacancy centre below, with each P1 centre treated as an explicit electron–$^{14}$N unit, give at 1 ppm (1.76 × $10^{17}$ $cm^{-3}$) and with the $^{13}$C bath included $T_2$ = 179 ± 22 µs·ppm. That is 12 % above the measurement, 0.8 combined standard deviations, with no adjustable parameter; ref. 19 obtains 188 µs·ppm for the same unit and field, with 24 configurations. With the outer spins frozen (static mean field), the same code gives 212 ± 39 µs·ppm, 33 % above the measurement, in line with earlier converged calculations of a static P1 bath that range from 236 to 286 µs·ppm.[20] Pair flip-flops shorten it by a factor 1.18 ± 0.15, in the direction ref. 19 also finds. Reducing the hyperfine interaction to a scalar detuning on the same positions (Methods) gives 161 ± 13 µs·ppm with pair flip-flops and 200 ± 23 µs·ppm frozen; the explicit unit and the scalar model differ by a factor 1.11 ± 0.13, so they are consistent. In that scalar model, letting each outer spin jump independently, rather than together with a resonant partner, gives 142 ± 20 µs·ppm.

With the $^{14}$N hyperfine switched off, the echo drops to 52 µs·ppm: the on-site hyperfine alone accounts for a factor 3.5 ± 0.3 in $T_2$, a reminder that a bath's internal frequency structure, not merely its density, sets the coherence. This work concerns $T_2$. $T_2^*$ carries no flip-flop dynamics; for the same baths it is collected in Supplementary Note 7, where the P1 value of 7.6 ± 0.7 µs·ppm is compared with the measured 9.6 ± 0.9 µs·ppm.

The calculation therefore reproduces the Hahn-echo time of a real spin bath within its statistical uncertainty. Everything that follows applies the same calculation, unchanged, to baths that have not been measured. The results are predictions.

### Bath frequency structure

Figure 2 shows the EPR transition spectrum of a single bath spin at 300 G for representative centres. Each line carries the probability that a randomly chosen defect presents it. The contrast is stark. $V^-$ has exactly one line, at the bare electron Zeeman frequency of 840.8 MHz. Three things keep it single: T_d symmetry forbids fine structure for $S$ = 3/2, the $^{12}$C host carries no nuclear spin, and at 50 ppm $^{13}$C fewer than one $V^-$ in a thousand has a $^{13}$C neighbour. Every $V^-$ is therefore resonant with every other one, and the resonant flip-flop fraction defined in Methods is P_res = 1. P1 splits into five lines spanning 228 MHz, through its own $^{14}$N hyperfine combined with the four Jahn–Teller axes; P_res = 1/4. The multi-vacancy centres split into four to twelve lines spanning 54–929 MHz, through the fine structure combined with the symmetry-equivalent site orientations; P_res falls between 0.097 and 0.250. Table 1 lists them all.

Two features of Table 1 control everything that follows. First, the number of lines does not track the magnitude of $D$. R8 has a large $D$ = 505 MHz, but its $C_{2v}$ tensor with $E$ = 0 and a ⟨110⟩ axis produces only four distinct frequencies. R7 has a similar $D$ = 580 MHz, yet its monoclinic tensor with $E$ = 33 MHz and tilted minor axes produces twelve. Second, the long chains have small $D$. The spin pair at the two ends of an 8V chain sits 1.13 nm apart, so KUL11 has $D$ = 53 MHz, and its lines separate by less than the width the bath itself imposes at high concentration. Larger is not automatically better.

**Concentration dependence**

Figure 3 gives $T_2$ for every centre at 0.01, 0.1, 1 and 10 ppm (Supplementary Tables 1–3). Concentrations are given in ppm of the carbon atom density, 1 ppm = $1.76 \times 10^{17}$ cm$^{-3}$. The echo is computed at 1 ppm and transferred to the other concentrations by the $1/n$ scaling (Methods). Direct calculations at 0.1 and 10 ppm reproduce that scaling for $V^-$ and R5, and for O1 to within 8 %, and show where it fails, for the chains R6, R10, R11 and KUL11 (below); the direct values are used for all of these. A single slope characterises every other bath. The enhancements quoted below are at 1 ppm unless stated.

At 1 ppm every multi-vacancy centre lands between 3.0 and 6.1 times the $V^-$ value: R8 2.97 < [V–V]$^0$ 3.52 < R5 3.82 ≈ KUL11 3.83 < R11 4.19 < R7a 5.18 < R7 5.27 < R6 5.61 ≈ O1 5.65 < R10 6.08, each with a statistical uncertainty of 8–14 %. That is not the order of $|D|$, which runs 795, 592, 580, 505, 465, 311, 182, 114, 76, 53 MHz for R5, R7a, R7, R8, [V–V]$^0$, O1, R6, R10, R11, KUL11. P_res sets the coarse order: 0.250 for R8, 0.187 for [V–V]$^0$, 0.125 for the chains and R7a, 0.097 for R7. Within the group at P_res = 0.125 a spread of 1.6 remains, and it is significant: R5 reaches only 0.68 ± 0.05 of O1. Along the chains the enhancement rises from 3.8 for R5 ($|D|$ = 795 MHz) to 5.6–6.1 for O1, R6 and R10, and drops again for R11 and KUL11, whose differently oriented sites lie close in frequency (below). The eigenstates of R5 carry the smallest moment along the field (root-mean-square ⟨S_z⟩ of 0.73, against 0.80–0.82 for the long chains), which weakens the static field through which the bath detunes its own nominally resonant pairs. R7a, with a moment nearly as small (0.76), nevertheless reaches 5.2, so this cannot be the whole explanation, and the low rank of R5 remains open.

P1 deserves a closer comparison, because P1 is what limits ordinary nitrogen-doped diamond. No single multi-vacancy species beats it, although a mixture of them can (below). P1 with its $^{14}$N hyperfine reaches an enhancement of 9.39 over $V^-$, above every multi-vacancy centre. P1 gains twice over. Its P_res is 0.250, comparable to the multi-vacancy centres, but it also carries $S = 1/2$ rather than $S = 1$, and the smaller moment is worth a further factor 2.7 on its own: a fictitious hyperfine-free P1 control already reaches 2.72 with P_res = 1. The two contributions separate cleanly: the $^{14}$N hyperfine multiplies P1′s $T_2$ by 9.39/2.72 = 3.46. Fine structure plus site orientation multiplies an $S = 1$ bath by a comparable 2.74 (next subsection). The two mechanisms are the same physics acting through different interactions.

Figure 4 replots the same information as a single slice at 1 ppm, ordering the centres by vacancy count. The step from the isolated vacancy $V^-$ to any multi-vacancy complex is the large one. Among the multi-vacancy centres the variation is smaller, a factor 2.0, and follows P_res at the coarse level.

Supplementary Tables 1–3 show two further points. First, the $^{13}$C bath at 50 ppm is negligible for $T_2$; it shortens every echo time by at most 0.3 %. Second, the enhancements stay flat in concentration for every centre except the chains R6, R10, R11 and KUL11. From 0.1 to 10 ppm the enhancement falls from 6.4 to 2.2 for R11, from 6.1 to 2.3 for KUL11, from 6.4 to 2.7 for R10 and from 6.5 to 3.4 for R6, while O1 loses only 6 %, from 5.7 to 5.4, and R5 does not change within its uncertainty (direct calculations, Methods). At 0.1 ppm the chains O1 to KUL11, which share P_res = 0.125, therefore come together at 5.7–6.5, and much of the spread among them at 1 ppm is this loss. The fall appears with a frozen outer bath as well, so the flip-flops of the outer bath are not its main cause. Three controls locate it (Supplementary Note 4). R5, with the largest $|D|$, and R5 with $D = 0$ both follow $1/n$, so a large $D$ alone does not break the scaling. Forcing every R11 or KUL11 defect onto a single orientation restores it: the ratio of $T_2 \cdot n$ at 10 ppm to that at 0.1 ppm goes from 0.30–0.37 to 0.99–1.00. Doubling the field at a fixed concentration of 1 ppm multiplies $T_2 \cdot n$ of R11 and KUL11 by 0.72–0.83, whereas a frozen single-orientation bath does not respond (1.00–1.01). The fall therefore needs sites of different orientation, and it depends on how far apart in frequency those sites lie. For a ⟨110⟩ chain with the field along [111] the first-order transition frequencies of differently oriented sites coincide, and only small shifts separate them: numerically about $0.225\, D^3/(\gamma B)^2$ in a third of the channels (47 kHz for KUL11 and 140 kHz for R11 at 300 G) and about $|E|$ in the rest (420 and 280 kHz). The bath coupling at the mean spacing, about 0.9, 9 and 92 kHz at 0.1, 1 and 10 ppm, grows towards these shifts. The weight of such near-resonant channels within 1 MHz, relative to the exactly resonant weight, ranks the fall across the chains: 0.004 for O1, 0.5 for R6 and 2.5–3 for R10, R11 and KUL11. This is a correspondence across five centres and not a quantitative model; R6 falls at 10 ppm although its main shift, 1.9 MHz, is some twenty times the typical coupling. The calculation also omits $g$ anisotropy and the inhomogeneous linewidth of real samples, either of which can change shifts of this size, so the onset concentration is a property of the model bath. The fine structure must nevertheless be large enough — not in absolute terms, but relative to the coupling it has to detune.

**What is responsible: a controlled decomposition**

A multi-vacancy centre differs from $V^-$ in three ways at once. It has $S = 1$ rather than 3/2, it has a non-zero fine-structure tensor, and that tensor points along one of several symmetry-equivalent directions. To separate the three, the same R5 bath was used throughout and only the assignment of the fine-structure tensor changed (Fig. 5, Table 2).

The step from $S = 3/2$ to $S = 1$ buys a factor 1.40 ± 0.08. That factor is trivial: a smaller spin carries a smaller magnetic moment. A 795 MHz fine-structure tensor, with every defect still on the same site, buys little: the next factor is 1.19 ± 0.07. Two identically oriented centres remain exactly resonant with each other on any one transition, because the fine structure shifts their common frequency without detuning them from one another. The small gain may come from the one exchange that the splitting closes even between identical sites, $|+1, -1\rangle \leftrightarrow |0, 0\rangle$, which costs an energy of order $2D$. The bath freezes only when the tensor may point along the twelve symmetry-equivalent ⟨110⟩ sites, and that step alone is worth 2.30 ± 0.20: most of the gain beyond the trivial spin factor (2.74 ± 0.24), or 83 % of it on a logarithmic scale. The uncertainties are statistical, from resampling the spatial configurations (Methods).

This is the central result, and it is the model of Fig. 1 made quantitative. The fine structure as such protects the $NV^-$ very little; the variety of fine-structure orientations does the work. A defect and its neighbour usually sit on differently oriented sites, and so fall out of resonance. Equivalently: P_res sets the scale, and the size of $D$ does not.

**A measured defect composition**

The calculation is finally applied to a composition that has actually been measured. Yamamoto et al. quantified the residual paramagnetic defects in nitrogen-implanted diamond by ensemble EPR, as a function of in-situ annealing temperature.[11] Their concentrations are not tabulated, so every figure quoted here was read off the semi-logarithmic plot in their Fig. 2(b), species by species and temperature by temperature, to an accuracy of roughly 30 %; the Supplementary Information lists what was extracted and the checks applied to it. After a 1000 °C anneal the inventory holds ⟨110⟩ multi-vacancy chains ($V_3$ variants and $V_4$, all $S = 1$) totalling 0.22 ppm, plus an isotropic broad signal with no fine structure at 0.74 ppm — 0.96 ppm in total. Each chain species received the published spin-Hamiltonian parameters of the centre it was identified with. The isotropic signal was treated as an $S = 1/2$ centre with no fine structure. Figure 6 and Table 3 set that composition against single-species baths of the same total spin density.

Three comparisons matter. Take first cases 5 and 6 of Table 3. The measured chain population, at its own measured density of 0.22 ppm, gives $T_2$ = 429 µs; the same density of $V^-$ gives 87.7 µs, a factor 4.9. That exceeds the 3.8 which R5 alone achieves in Fig. 3, and case 7 explains why. A bath of R5 alone at the same density gives 336 µs, so mixing the measured chain species multiplies $T_2$ by a further 1.28 ± 0.14. Two defects of different species have different $D$, and that detunes them as surely as two defects of the same species on different sites; the mixture's P_res is 0.090 against 0.125 for R5 alone. A heterogeneous bath is therefore not equivalent to a homogeneous one of the same density, in line with what defect-resolved simulations of mixed nitrogen-, vacancy- and hydrogen-related baths have found.[17]

Take next cases 1 and 2. With the isotropic signal included at its measured density, the whole sample gives 88.4 µs against 20.0 µs for an equal density of $V^-$, a factor 4.4. The benefit survives intact.

Cases 1 and 4 isolate what the isotropic signal costs. If more multi-vacancy chains replace it, at a fixed total spin density of 0.96 ppm, $T_2$ moves only from 88.4 to 97.5 µs (a ratio of 1.10 ± 0.13), although P_res drops from 0.601 to 0.090; at 900 °C, where P_res drops from 0.294 to 0.063, the ratio is 0.94 ± 0.13. The isotropic species carries 77 % of the spins in this sample, yet within the statistical uncertainty it is no more damaging than the chains it would be replaced by. Its spins flip-flop freely among themselves, but each carries $S = 1/2$, and the smaller moment offsets the free flip-flops. P_res compares baths of one spin quantum number; across spin quantum numbers the moment counts as well. A P1 bath of the same total density would still give more, 187 µs (case 3), as expected from P1′s smaller spin and its own hyperfine structure.

The same exercise at 800 and 900 °C separates the two things an anneal does: it lowers the concentration, and it changes the composition (Fig. 7, Supplementary Table 4). The total defect density falls by a factor 5.2 from 800 to 1000 °C, from 4.96 to 0.96 ppm, which on its own would lengthen $T_2$ by the same factor. But the composition degrades. The simulated $T_2$ improves by only 1.84 (48.0 ⟶ 88.4 µs), so the product $T_2 \cdot n$ falls from 238 to 85 µs·ppm, a factor 2.8 ± 0.4. The isotropic signal is not responsible: a bath of multi-vacancy chains only, at each temperature's total density, loses nearly as much (242 ⟶ 93 µs·ppm, a factor 2.6 ± 0.3). What degrades is the variety of the chain population. The anneal removes the longer chains first: six chain species at 800 °C give way to three at 1000 °C, R5 rises from 47 % to 84 % of the chains, and the P_res of the chain population rises from 0.047 to 0.090. The bath loses the mixture effect of cases 5 and 7.

Before that loss, the mixture carries the bath past P1. At equal spin density the 800 °C composition gives $T_2 \cdot n$ = 238 µs·ppm and its chains alone 242 µs·ppm, 1.33 ± 0.17 and 1.35 ± 0.21 times the 179 µs·ppm of a P1 bath (Fig. 7b). These lie 1.9 and 1.7 standard deviations above parity, with 95 % bootstrap intervals of 1.07–1.67 and 1.07–1.80; the uncertainties count the sampling of configurations only, not the reading of the composition (Discussion). Most of those spins carry $S = 1$, twice the moment of P1, but six chain species at P_res = 0.047 outweigh that, against P_res = 0.250 for P1. The advantage goes as the variety goes: 0.78 ± 0.10 at 900 °C and 0.47 ± 0.08 at 1000 °C for the measured compositions.

This is a statement about the damage done per spin, not about the sample: at its total density of 4.96 ppm the 800 °C composition has $T_2$ = 48 μs.

The series stops at 1000 °C, although the measurement extends to 1200 °C. Above ~1100 °C the chains have annealed out and two other signals take over, labelled H and R12 in that work.[11] Neither can be put into this calculation. H is assigned only tentatively, as a multi-vacancy cluster lacking the $C_{2v}$ symmetry of the chains, and no fine-structure tensor has been published for it. R12 ($C_{3v}$, $S$ = 1) does have published parameters, but its spectrum overlaps the isotropic signal O, so the two concentrations cannot be read separately. Since P_res is set by the fine-structure tensor and the site multiplicity, extending the series upward would mean inventing the very quantity the argument turns on.

---

## Discussion

The model of Fig. 1 survives the test, with a calculation that first reproduces the measured P1 bath. The model asserted two things: that a bath blocks its own flip-flops when its members present different EPR transition frequencies, and that aggregation supplies exactly that spread. The calculations bear both out. The enhancement follows P_res and not the size of the splitting (Fig. 4 and Table 1). Holding the fine structure fixed but forcing every defect onto one orientation destroys most of the benefit beyond the spin factor: a single orientation gives 1.19, the twelve real orientations 2.30 (Fig. 5 and Table 2). And mixing species, which widens the frequency spread further still, adds another factor 1.28 (Table 3). The rest of this section takes the model as established and asks what it implies, and where it stops.

$T_2$ is not monotonic in $D$. Ordered by enhancement at 1 ppm, the multi-vacancy centres run R8 (2.97) < $[V\text{–}V]^0$ (3.52) < R5 (3.82) ≈ KUL11 (3.83) < R11 (4.19) < R7a (5.18) < R7 (5.27) < R6 (5.61) ≈ O1 (5.65) < R10 (6.08). Ordered by $|D|$ they run 505, 465, 795, 53, 76, 592, 580, 182, 311, 114 MHz — an unrelated sequence. The centre with the largest $|D|$, R5 (795 MHz), sits in the lower half of the list, while the chains O1, R6 and R10, with $|D|$ of 114–311 MHz, head it. R8 makes the point sharply. Its $D$ = 505 MHz is among the largest, but its axial $C_{2v}$ tensor ($E$ = 0) along ⟨110⟩ generates only four distinct frequencies, so a quarter of all pairs stay resonant and it shows the smallest enhancement of the set. R7 has essentially the same $D$, but its monoclinic tensor with $E$ = 33 MHz and tilted minor axes generates twelve frequencies, and it reaches the lowest P_res of any centre considered; it nevertheless does not exceed O1, R6 and R10, so once the bath's own static field is included P_res is not the only factor. What matters is how many distinct transition frequencies the bath presents. The site multiplicity and the rhombicity $E$ set that number; $D$ affects it only indirectly. $D$ acts as a threshold rather than a magnitude. It must be large enough that differently oriented sites stay detuned against the bath–bath coupling, which is where R6, R10, R11 and KUL11 fail at high concentration. Beyond that threshold, making it larger buys nothing: R5 gains less than O1 and R10 at the same P_res.

The ratio $T_2/T_2^*$ offers a direct experimental signature that needs no absolute concentration calibration, the largest single uncertainty in this kind of comparison, because a bath that blocks its own flip-flops shows a large ratio. A bath with no frequency structure gives 5–8 and the multi-vacancy baths 11–22 (Supplementary Note 7).

The two experiments that motivated this work can now be read again. Neither result is in doubt. What the calculation changes is what they can be attributed to. Yamamoto et al. obtain a long $T_2$ after a 1000 °C anneal and credit the removal of multi-vacancy clusters.[11] In their composition, reproduced here, those clusters are the benign part of the bath: an equal density of isolated $V^-$ in their place shortens $T_2$ by 4.4. The anneal helps because it removes spins, not because it removes clusters. It helps less than the fall in spin count would suggest, because it also removes the longer chains and with them the variety of the chain population. That is a testable prediction: the product of $T_2$ and the total paramagnetic spin density should fall between 800 and 1000 °C, by a factor of about 2.8.

Herbschleb et al. obtain the record room-temperature $T_2$ = 2.4 ms in phosphorus-doped n-type diamond and credit charging of the vacancies, which suppresses the formation of paramagnetic vacancy complexes.[14] Here the reinterpretation is sharper, because the charge state does two opposite things at once. Suppose that suppressing aggregation leaves the constituent vacancies behind in the negative charge state. The bath then gets worse, not better. $V^-$ is the single most damaging centre in Table 1, with P_res = 1 and $S$ = 3/2, and $n$ of them carry $n$ times the spin of the one chain they would have formed. If one $n$-vacancy chain breaks back into $n$ isolated $V^-$, the bath loses a factor 3.0–6.1 from its frequency structure, and a further factor from the $n$-fold rise in centre count. The benefit reported in ref. 14 therefore cannot come from suppressing complex formation as such. It must come from the vacancies leaving altogether — by recombination with interstitials, by out-diffusion, or by never being retained. The charge state acts as the lever that keeps them mobile and unbound, not as a way of turning a harmful species into a harmless one.

The distinction matters for process design, because it changes the sign of the target. "Suppress multi-vacancy formation" and "reduce the total paramagnetic spin count" coincide only when the vacancies actually leave. Where they do not — where charging merely prevents them from binding — the second instruction is right and the first is counterproductive.

One boundary condition keeps this claim honest. Where the vacancy is neutral, $V^0$ is diamagnetic, and aggregation then creates paramagnetic $S = 1$ centres where there were none. In that regime suppressing aggregation is unambiguously good, and the conventional reading holds. The charge state of the vacancy, not the aggregation, fixes the sign of the effect. To tell the two cases apart, an experiment need only establish the vacancy charge state by EPR alongside the coherence measurement.

The implications for processing follow directly. Two samples with the same total paramagnetic defect density can differ in $T_2$ by a factor of several, depending on what those defects are. Vacancy aggregation pushes in the favourable direction on both counts. It cuts the number of paramagnetic centres, since $n$ negatively charged vacancies that would have carried $n$ spins of $S = 3/2$ become one centre of $S = 1$. It also gives that surviving centre a fine structure whose orientation varies from site to site. What cancels the benefit is a species that is both structureless and high-spin, $V^-$ itself. The structureless $S = 1/2$ signal that survives annealing[11] costs no more than the chains at equal density, because its smaller moment offsets its free flip-flops. The calculation therefore points at the total paramagnetic spin count and at the charge state of the vacancies, not at suppressing vacancy aggregation.

Six limitations of the calculation deserve attention. First, the outer bath is treated as pair flip-flops with rates from the dipolar Hamiltonian, propagated as a time-dependent field on each cluster.[19] That treatment reproduces the measured P1 $T_2$ (Results), but it was validated on that one bath; ratios to $V^-$ carry a statistical uncertainty of 8–14 %, and the pair-flip-flop range and the number of trajectories add a dependence of up to 15 % that is not included in that figure (Methods). The concentration dependence is computed directly for $V^-$, R5 and the chains O1 to KUL11; for the other centres it rests on the $1/n$ scaling, which their lack of near-resonant channels supports but which was not computed directly. All conclusions rest on ratios computed under identical conditions, which cancels errors common to the baths but not errors that differ between them. Second, the expansion is truncated at order 2, and convergence in the order is not established here (Methods). $V^-$ sets the denominator of every enhancement ratio, and the published convergence tests cited in Methods do not cover it: it has a single line and $S = 3/2$. A truncation error that differs between $V^-$ and the multi-vacancy baths would change those ratios; this work does not quantify it, and extending those tests to $S = 3/2$ remains an open task. Third, the calculation treats pure dephasing only and omits the bath spins' own longitudinal relaxation. This matters for the neutral divacancy R4/W6, whose EPR linewidth broadens sharply above ~50 K through an Orbach process involving an excited state 20(1) meV above the ground state;[16] its entry in Table 1 is therefore closer to a low-temperature statement. The vacancy chains and the clusters R7a and R8 show no temperature dependence of $\mathbf{D}$ over the range studied,[12] so the treatment applies more directly to them. Fourth, the symmetry-equivalent site orientations are assumed equally populated. Stress-induced alignment during growth or irradiation would reduce the effective site multiplicity and move the bath towards case C of Table 2, that is, towards losing the orientation factor; stress-dependent EPR would test this. Fifth, $NV^-$ spin-lattice relaxation is omitted. At room temperature $2T_1 \approx 12$ ms caps any $T_2$ quoted here, and that cap binds at about 0.01 ppm for the longest-lived multi-vacancy centres. Sixth, the measured composition of the Results was digitised from a published semi-logarithmic figure rather than read from a table, to about 30 %. That uncertainty enters the absolute times of Figs 6 and 7 roughly linearly, since $T_2$ scales as $1/n$. It does not enter the comparisons drawn from them, because those hold the total spin density fixed and change only what the bath is made of; what the reading error does affect is the species fractions, and through them P_res, so the margin over P1 at 800 °C carries an uncertainty beyond its statistical one.

In summary, a cluster-correlation expansion whose outer bath keeps flip-flopping reproduces the measured Hahn-echo time of the P1 bath within its statistical uncertainty, and applied unchanged to multi-vacancy baths it predicts that the aggregation of negatively charged vacancies into multi-vacancy complexes does not shorten $NV^-$ spin coherence. At equal paramagnetic spin density it lengthens the Hahn-echo coherence time by a factor 3.0–6.1 at 1 ppm and 2.2–6.5 from 0.01 to 10 ppm, because the complexes present many distinct EPR transition frequencies and so block one another's flip-flops. The symmetry-equivalent site orientations of each centre supply most of that gain beyond the spin factor. In annealed diamond neither the multi-vacancy chains nor the structureless isotropic species that survives annealing is the problem; the coherence follows the total spin count and the variety of the chain population. The quantity that decides the outcome, the fraction of bath pairs sharing a transition frequency, can be evaluated in the same way for the high-spin multi-vacancy complexes of silicon and silicon carbide.

## Methods

### Spin Hamiltonian and bath construction

The central spin is the $NV^-$ ground-state triplet: $S = 1$ with $D = 2.87$ GHz, quantised along its [111] symmetry axis. The static field $B = 300$ G lies parallel to that axis. The qubit is the $|m_s = 0\rangle \leftrightarrow |m_s = -1\rangle$ transition. All calculations run in the NV frame, where $z \parallel [111]$.

Each bath spin $i$ carries

$$H_i = \mathbf{I}_i \cdot \mathbf{P}_i \cdot \mathbf{I}_i + \gamma_i \mathbf{B} \cdot \mathbf{I}_i + \delta_i I_{z,i},$$

and pairs interact through the point dipolar tensor. $\mathbf{P}_i$ is the fine-structure (zero-field splitting) tensor of that centre; $\delta_i$ is an on-site hyperfine detuning, used only for the scalar P1 model below. The spin-Hamiltonian parameters come from the literature without modification and are collected in Table 1. For each centre, the traceless fine-structure tensor is built from its published principal values, oriented along the crystallographic axis reported for that centre, then rotated into the NV frame. Every bath spin receives one of the symmetry-equivalent site orientations, drawn uniformly and independently: twelve for the ⟨110⟩-axial chain and cluster centres, and twelve for the $C_{2h}$ divacancy. This site randomness is not a refinement. As the Results show, it is the dominant part of the effect.

Concentrations are given in ppm of the carbon atom density of diamond (1 ppm = $1.76 \times 10^{17}$ $cm^{-3}$). Bath positions are drawn uniformly in a sphere at the requested number density $n$, with a Poisson number of spins. The effect under study is a property of the pair statistics of the bath, so the cutoffs scale with concentration as $n^{-1/3}$. Every concentration then keeps the same number of bath spins inside $r_{bath}$ and the same number of pairs inside $r_{dipole}$. Without that scaling, the concentration series would mix a physical trend with a convergence trend. The Hahn echo uses a cluster cutoff of several mean spacings (see CCE protocol): $r_{dipole} = 4\, n^{-1/3}$ and $r_{bath} = 5.5\, n^{-1/3}$, about 700 spins, that is $(r_{bath}, r_{dipole}) = (981, 714)$ Å at 1 ppm.

### The P1 centre

P1 requires care, because its frequency spread is the natural yardstick for the multi-vacancy spread. A P1 centre is an $S = 1/2$ electron coupled to its own $^{14}N$ nucleus ($I = 1$) through an axially symmetric hyperfine tensor, with a quadrupole term on the nucleus. That tensor's axis is the Jahn–Teller distorted N–C bond, one of the four ⟨111⟩ directions, drawn at random for each centre. The echo treats every centre as this explicit six-level unit, as in ref. 19: a sampled bath state gives each unit one of its six eigenstates with equal probability, and a pair cluster is two such units. In the high-field limit the unit reduces to a scalar detuning $\delta = m_I A_{zz}(\theta)$, with $A_{zz}(\theta) = \sqrt{A_\parallel^2 \cos^2\theta + A_\perp^2 \sin^2\theta}$, $A_\parallel = 114.03$ MHz, $A_\perp = 81.31$ MHz, and $\theta$ the angle between the centre's Jahn–Teller axis and the field.[21,22] That form gives the line count and $P_{res}$ of Table 1. With $\mathbf{B} \parallel [111]$, one of the four axes lies parallel to the field and three make $\cos\theta = -1/3$. $A_{zz}$ therefore takes two values, 114.03 and 85.57 MHz, and the five distinct P1 transition frequencies $\gamma_e B + \delta$ carry weights 1/12, 1/4, 1/3, 1/4, 1/12.

The high-field reduction does not hold at arbitrary field, so exact diagonalisation of the coupled electron–$^{14}N$ Hamiltonian, including the quadrupole term, verified it at the field used here. The expectation value of the electron spin along the field, $\langle S_B \rangle$, spans 0.035–0.500 at 25 G, 0.434–0.500 at 100 G and only 0.493–0.500 at 300 G. At 25 G the electron and its nucleus hybridise strongly and the reduction fails; at the field of this work the electron is cleanly quantised. The exact transition frequencies differ from the scalar ones by at most 11 MHz. That second-order shift is common to all members of a degeneracy class, so the five-way degeneracy structure — and hence the resonant fraction — is unchanged, and every mismatch between classes stays four orders of magnitude above the dipolar coupling. The scalar model, run on the same positions, is kept as a reference for the echo (Results). A fictitious "P1 without hyperfine" bath, a bare $S = 1/2$ electron with isotropic $g$, is also run. It is unphysical, but it isolates how much of P1's advantage over $V^-$ comes from the $^{14}N$ hyperfine rather than from the smaller spin quantum number.

### Resonant flip-flop fraction

$P_{res}$ captures in one number how strongly a bath blocks its own flip-flops. It is the weighted probability that two bath spins, drawn independently with their site orientations and nuclear states — and, for a multi-species bath, with their species drawn in proportion to concentration — present EPR transition frequencies agreeing to within a threshold. A bath with a single transition frequency gives $P_{res} = 1$. A bath spreading its weight over $k$ equally likely frequencies gives $P_{res} = 1/k$. Applied to $^{14}N$ P1, the definition returns exactly 1/4, which matches a count of resonant channels by hand: the five transition frequencies carry weights 1/12, 1/4, 1/3, 1/4 and 1/12, and the sum of their squares is exactly 1/4.

The threshold must sit above the bath–bath dipolar coupling ($9 \times 10^{-5}$ to $9 \times 10^{-2}$ MHz over the concentration range studied) and below the smallest structural splitting the bath presents (47 kHz for KUL11 and 137 kHz for R11, from the nearly degenerate site orientations of the long vacancy chains). A value of 10 kHz is used, and every value between 1 and 40 kHz gives the same P_res for every centre in Table 1. No single threshold meets both conditions at 10 ppm, where the coupling exceeds the smallest of those splittings; P_res counts exact resonance only, and the long chains lose their advantage at high concentration (Results).

### CCE protocol

Every coherence function behind the figures and tables was computed with dynacce (version 0.1.0.dev0), a cluster-correlation-expansion code with a dynamical mean-field treatment of the bath developed by the author.[19] The bath configurations were generated with the author's own scripts, using PyCCE[23] data structures; the $^{13}$C positions were drawn with PyCCE's `random_bath`. The $NV^-$ zero-field splitting far exceeds every coupling in the problem, so the conventional (pure-dephasing) CCE applies. The Hahn echo is computed at CCE order 2, the order at which bath–bath flip-flops enter. Ensemble averages use 12 spatial configurations, each with four sampled bath states (below).

A $^{13}$C bath at 50 ppm is computed separately on the diamond lattice and multiplied in. That level is not a generic choice. It is the residual $^{13}$C concentration of the $^{12}$C-enriched high-pressure high-temperature crystals used in the author's own $NV^-$ dephasing measurements,[24] so the nuclear bath modelled here is the one that accompanies the material this work is meant to inform. On its own it gives $T_2 \approx 0.20$ s at this field, so it never limits $T_2$ here.

Coherence times are the 1/e points of the resulting decays. No lifetime anywhere in this work comes from a fit. Conventional CCE evaluates a ratio of cluster to subcluster contributions independently at each time point, so isolated points of the echo can jump above unity when a subcluster contribution passes through zero. A Hahn-echo envelope in this bath has no revival mechanism, so the running minimum of each echo decay is taken before the 1/e point is read off.

The echo calculation must carry the mean field that the spins outside each cluster exert on the spins inside it. A fully mixed bath has no such field, flip-flops inside a cluster go undetuned, and spectral diffusion runs too fast. The bath state is therefore sampled as a pure product state: each spin is assigned one eigenstate of its own single-spin Hamiltonian, drawn with equal probability (the room-temperature bath), and dynacce adds the mean field of the outer spins. That field is not frozen. The outer spins flip-flop in pairs: each pair within 4 mean spacings makes a joint, energy-conserving transition at a golden-rule rate set by its dipolar coupling and by its instantaneous detuning, which includes the mean field of the rest of the bath, so that a pair whose transition frequencies differ by more than the coupling hardly moves.[19] Each sampled bath state starts one stochastic trajectory of these flip-flops. Every single-spin cluster, and every two-spin cluster that responds appreciably to the field, is propagated in the resulting time-dependent mean field; the remaining clusters keep their static-mean-field contribution. Pair clusters are propagated exactly in every bath (Supplementary Note 8). The complex coherence is averaged over the four trajectories per configuration. Freezing the outer spins (static mean field) or letting each of them jump independently of a partner gives the comparison values of the P1 benchmark; Supplementary Note 6 lists both for every bath. With the mean field, convergence also requires a cluster cutoff of several mean spacings. Supplementary Note 4 shows that r_dipole = 3, 4 and 5 $n^{(-1/3)}$ give the same $T_2$ for $V^-$, R5, $[V–V]^0$ and P1 within the statistical uncertainty of the test (the largest difference is 1.40 ± 0.28, for $[V–V]^0$ at 5 mean spacings), whereas the fully mixed bath with r_dipole = 1.2 $n^{(-1/3)}$ underestimates $T_2$ by factors of 1.4 ($[V–V]^0$) to 2.2 (P1); the same behaviour was found for a static P1 bath by Park et al.[20] The pair-flip-flop range (3, 4 and 5 mean spacings) and the number of trajectories (4 and 8) change the $T_2$ of $V^-$, P1 (scalar model), R5 and the 1000 °C baths by up to 15 %, about the size of the statistical uncertainty and with no monotonic trend in the range; the explicit P1 unit was not part of this check (Supplementary Note 4). In the production runs the statistical uncertainty of a $T_2$ ratio to $V^-$, estimated by resampling the configurations (which are shared between baths), is 8–14 %. Because only the dipolar coupling scales with concentration, $T_2 n$ is independent of $n$ whenever the bath's internal splittings far exceed that coupling. The echo is therefore computed at 1 ppm and transferred to other concentrations as $L_n(t) = L(tn/1\ \text{ppm})$. For O1, R6, R10, R11 and KUL11, whose smallest splittings approach the coupling at high concentration, the echo is computed directly at 0.1 and 10 ppm as well, with $V^-$, R5 and R5 at $D = 0$ as controls, and 0.01 ppm is obtained from 0.1 ppm by the same scaling. The measured compositions of Table 3 and Fig. 7 are computed at 1 ppm with the measured species fractions and transferred to their measured total density in the same way. The species whose scaling fails (R6, R10, R11) are at most 0.43 ppm in those compositions, so the transfer should hold, but it was not computed directly. Order 3 could not serve as a convergence test. CCE beyond order 2 is known to turn unphysical ($|L| > 1$) in dilute dipolar electron baths, when a cluster holds only part of a strongly interacting group of spins[25] or when a lower-order cluster factor passes near zero and enters a higher-order term as a denominator.[26] Here, in a fully mixed test bath, CCE3 departs from CCE2 near $T_2$ and then diverges (Supplementary Note 4), so convergence in the order is not established in this work. The support for CCE2 is external. It reproduces the measured P1 $T_2$ (Results). A CCE that averages over bath states inside the expansion finds cluster corrections up to six spins slight for resonant spin-1/2 in three dimensions, and the same with four spins for P1;[26] a partitioned CCE finds P1 baths converged at this order, attributing that to the hyperfine substructure suppressing flip-flops, whereas spin-1/2 baths without it, in quasi-two-dimensional layers, need larger partitions.[25] These tests do not cover a bath of $S = 3/2$, so the $T_2$ of $V^-$, and the ratios taken to it, may carry a truncation error that this work does not quantify.

---

## Data availability

The data supporting the findings of this study — the simulated coherence decays behind every figure and table (Hahn echo and free induction decay, every centre, every concentration and every scenario) and the coherence times extracted from them — are available from the corresponding author on reasonable request. All input parameters of the simulations are published electron-paramagnetic-resonance spin-Hamiltonian parameters and are given in full in Table 1 and in the Supplementary Information, so the calculations can be reproduced from the paper alone.

## Code availability

The simulation code — the dynacce package (version 0.1.0.dev0) used for every coherence calculation, the EPR spin-Hamiltonian database, the bath-construction and site-orientation routines and the analysis scripts — is available from the corresponding author on reasonable request. PyCCE[23] is openly available.

## Acknowledgements

The author gratefully acknowledges the support of the Japan Society for the Promotion of Science (JSPS) KAKENHI Grant Number 25K17636.

## Author contributions

C.S. conceived the study, performed the simulations, analysed the results and wrote the manuscript.

## Funding

This work was supported by the Japan Society for the Promotion of Science (JSPS) KAKENHI Grant Number 25K17636 (C.S.).

## Competing interests

The author declares no competing interests.

---

---

## Figure legends

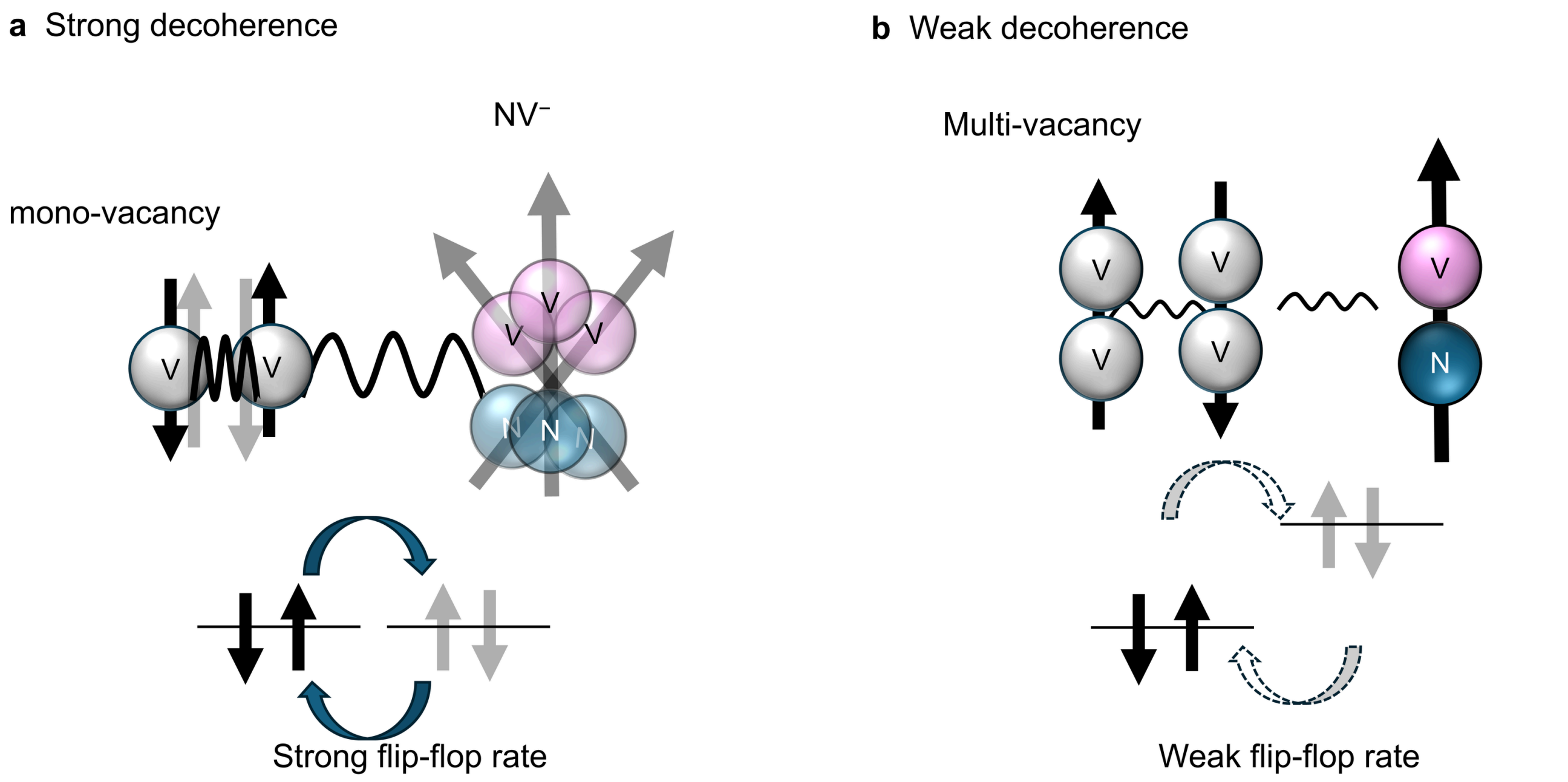

**Fig. 1 | Proposed model for the flip-flop bottleneck.** The picture this work sets out to test. **a**, A bath of isolated vacancies. Every $V^-$ presents the same EPR transition frequency, so any neighbouring pair is resonant, flip-flops proceed freely, and the dipolar field they produce fluctuates quickly at the $NV^-$. **b**, A bath of multi-vacancy complexes. Each complex carries a fine-structure splitting whose principal axis points along one of several symmetry-equivalent crystallographic directions, so two neighbouring defects generally sit at different transition frequencies. The pair is then off resonance and the flip-flop is suppressed. The level diagrams below each panel show the two bath spins involved: matched in **a**, detuned in **b**.

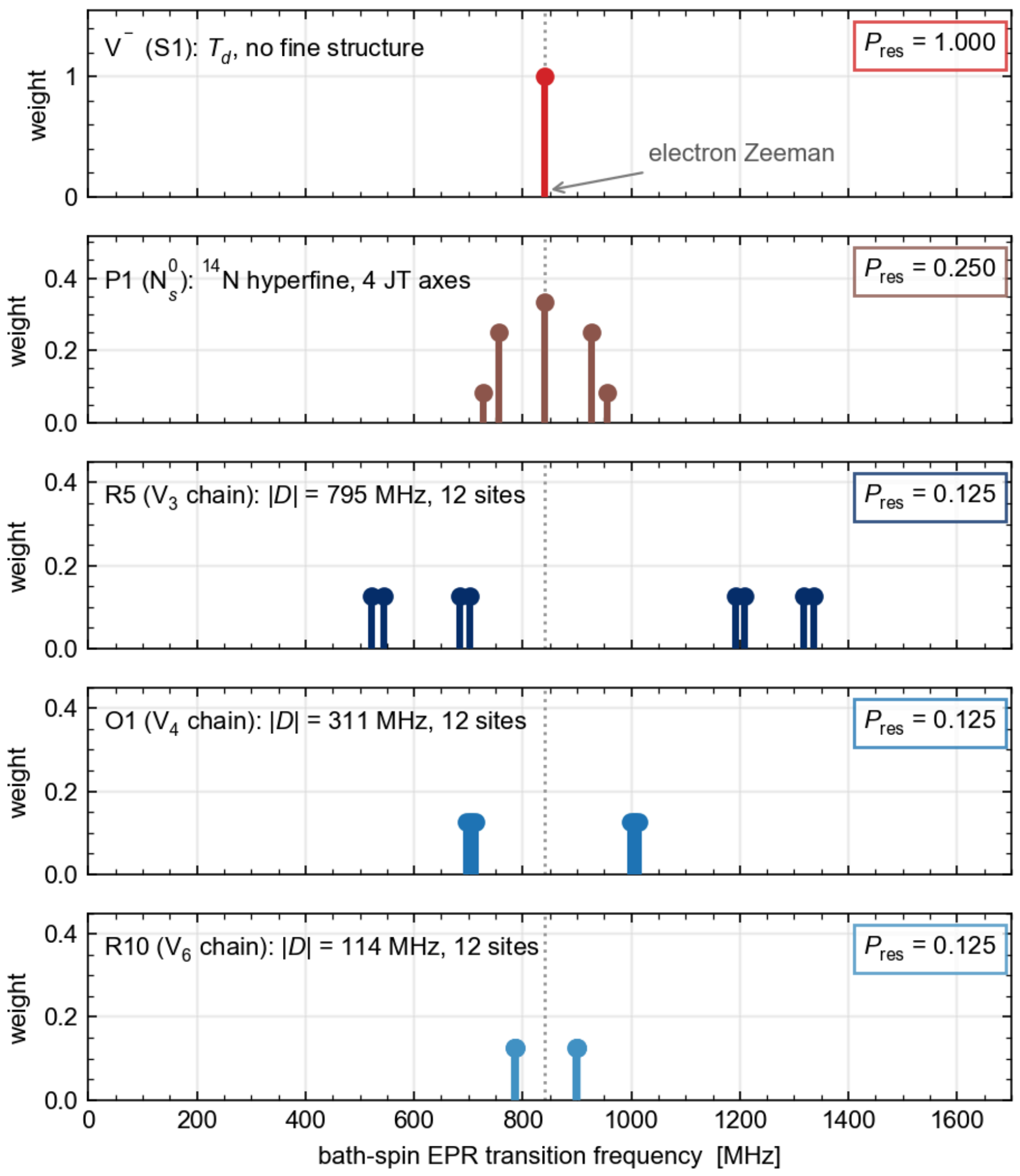

**Fig. 2 | Bath transition spectra and the resonant flip-flop fraction.** EPR transition spectrum of a single bath spin at $B$ = 300 G along [111], each line weighted by the probability that a randomly chosen defect presents it; the dotted vertical line marks the bare electron Zeeman frequency, indicated by the arrow. $V^-$ has a single line, so every pair of $V^-$ is resonant and P_res = 1. P1 splits into five lines through its $^{14}$N hyperfine combined with the four Jahn–Teller axes, and the multi-vacancy chains into eight through their fine structure combined with the twelve symmetry-equivalent site orientations. P_res is quoted in each panel.

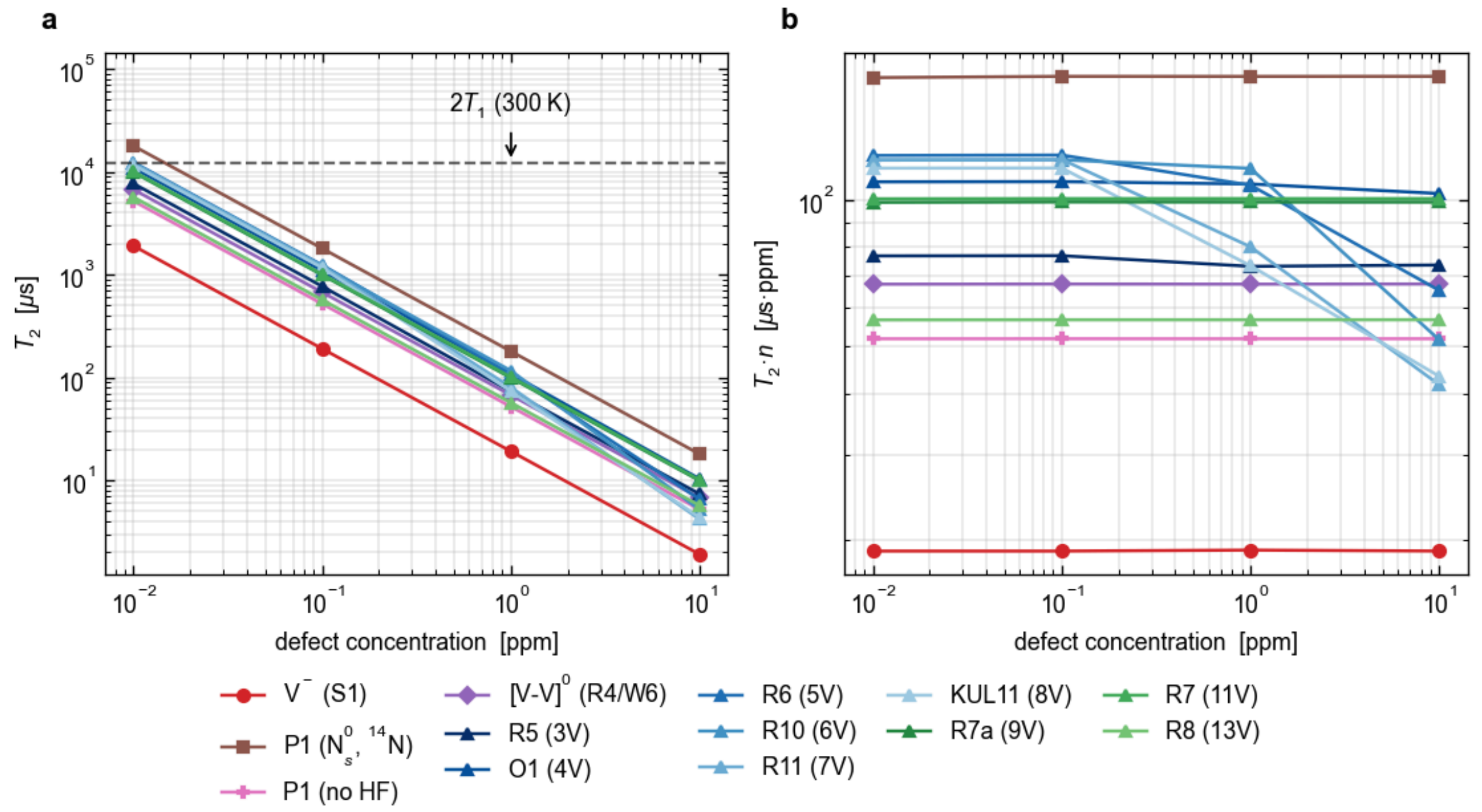

**Fig. 3 | Concentration dependence of $T_2$.** **a**, $T_2$ (Hahn echo, pair flip-flops of the outer bath) versus defect concentration for every centre considered; the dashed line is $2T_1$ at 300 K. **b**, The same $T_2$ data with the $1/n$ trend divided out, so that a bath following $1/n$ becomes a horizontal line whose height is its slope; the ordering of the baths, compressed in **a**, is then legible. The fall of R6, R10, R11 and KUL11 towards 10 ppm needs sites of different orientation and depends on their small frequency separation (Results).

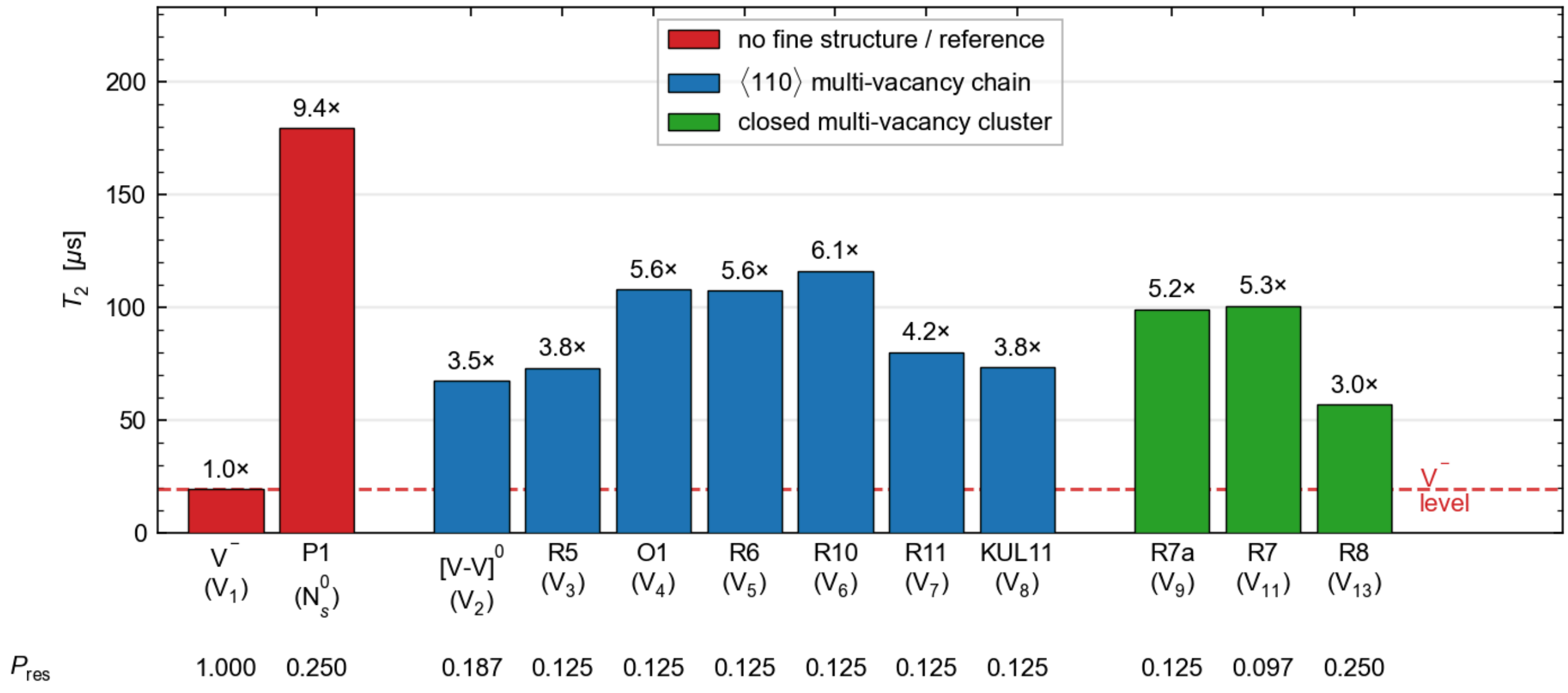

**Fig. 4 | $T_2$ by defect species at a fixed concentration.** The data of Fig. 3 taken as a single slice at 1 ppm and plotted against defect species, ordered by the number of vacancies in the complex. The enhancement over V⁻ is printed above each bar, the resonant flip-flop fraction P_res below each species label, and the dashed line marks the V⁻ level. The step from the isolated vacancy to any multi-vacancy complex is the large one.

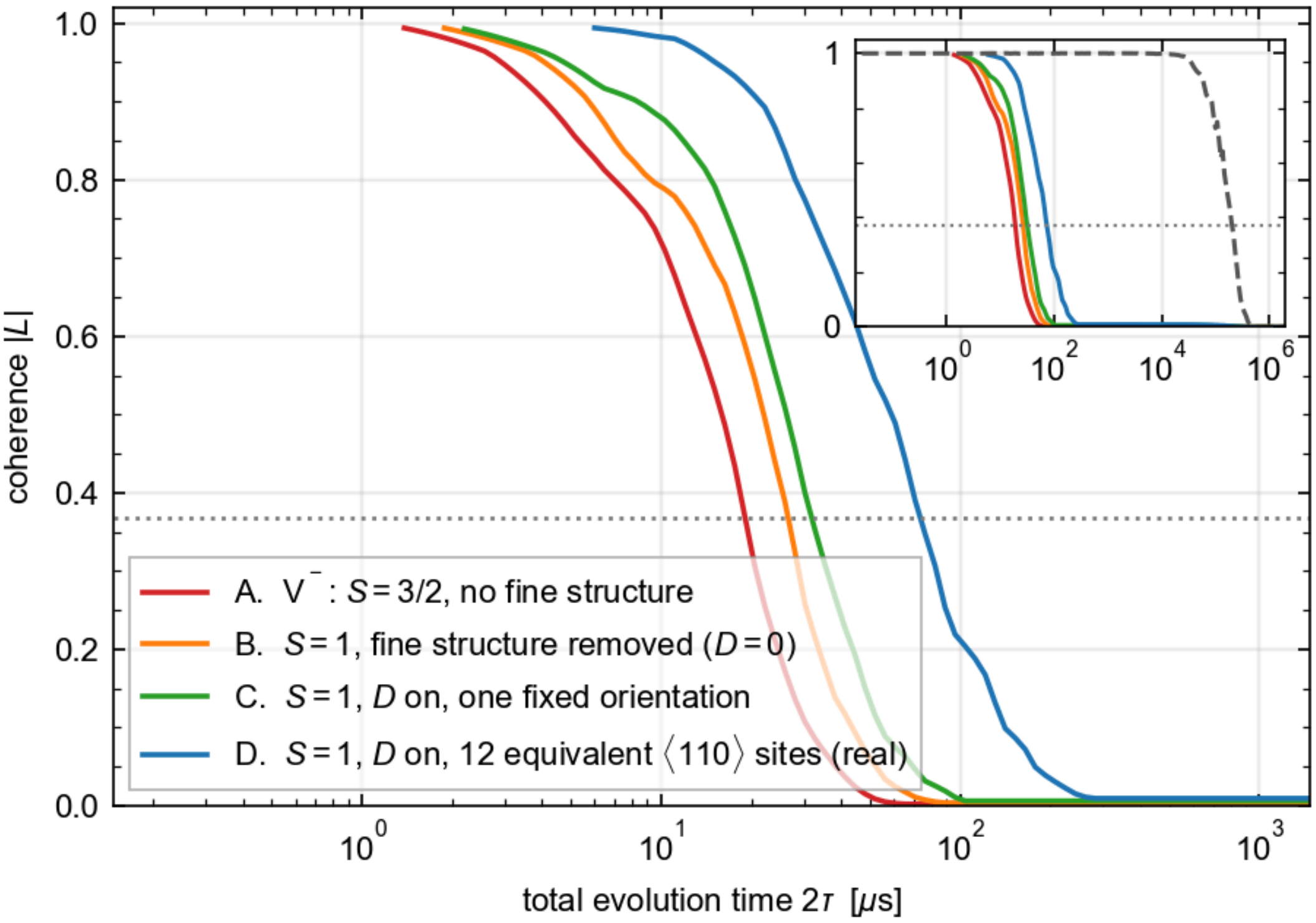

**Fig. 5 | What lengthens $T_2$: controlled decomposition at 1 ppm.** Only the assignment of the fine-structure tensor differs between the four curves; the positions, concentration, spin quantum number (except in A) and all cutoffs are identical. The dotted horizontal line marks 1/e. Inset: the same curves over the full time range, together with the $^{13}C$-only reference (dashed), showing that the nuclear bath is three orders of magnitude away from limiting anything here.

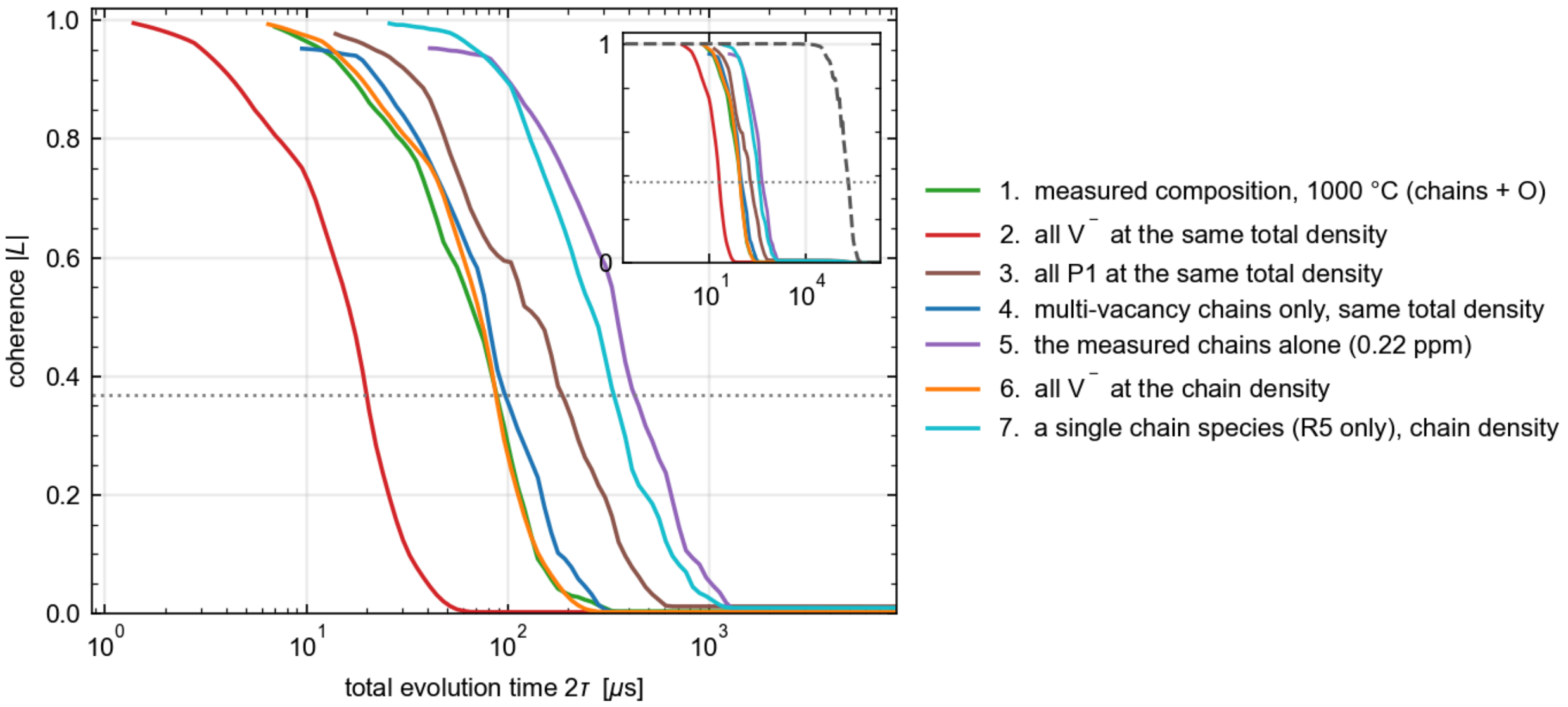

**Fig. 6 | $NV^-$ spin coherence in a measured defect composition.** The composition determined by ensemble EPR after a 1000 °C anneal, read from Fig. 2(b) of ref. 11 to about 30 %, against single-species baths of the same total spin density. Inset: the full time range with the $^{13}C$-only reference (dashed).

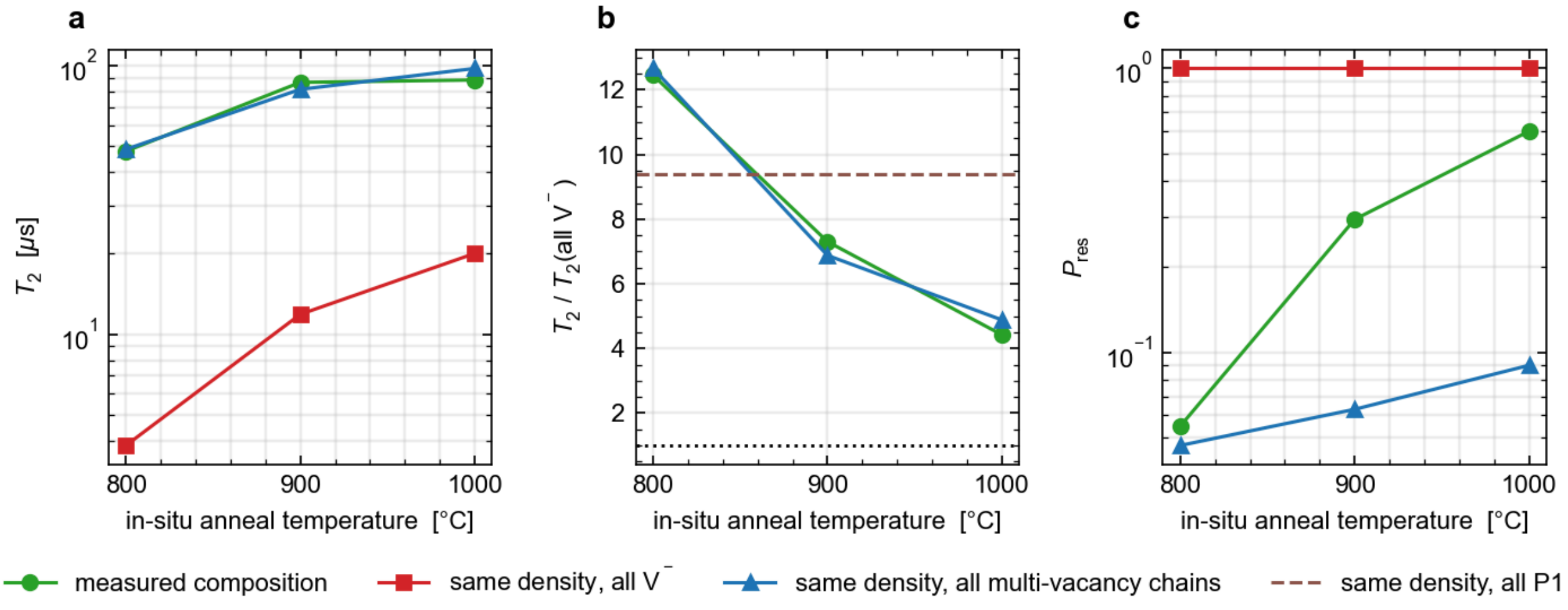

**Fig. 7 | Annealing series, 800–1000 °C.** The same calculation across the annealing series of ref. 11, with the composition at each temperature read from Fig. 2(b) of that work to about 30 %. **a**, $T_2$ versus anneal temperature for the measured composition, for an all-$V^-$ bath of the same total density, and for an all-multi-vacancy bath of the same total density. **b**, The ratio of the measured composition and of the all-multi-vacancy bath to the all-$V^-$ case, i.e. the gain that comes from the bath's frequency structure rather than from its concentration; the dashed line marks a P1 bath of the same density, which the 800 °C baths exceed. **c**, The resonant flip-flop fraction of each. That of the measured composition rises by an order of magnitude as the chains anneal out faster than the isotropic signal, yet $T_2$ follows the all-multi-vacancy bath, whose chain population loses its variety (Results).

## Tables

**Table 1 | EPR spin-Hamiltonian parameters and bath frequency structure at $B$ = 300 G along [111].** $|D|$ and $E$ are the conventional fine-structure parameters, "site orient." the number of symmetry-equivalent orientations the centre can take, "EPR lines" the number of distinct transition frequencies the centre presents, "spread" the range those lines span, and P_res the resonant flip-flop fraction defined in Methods.

| centre | structure | $S$ | $n_V$ | $\|D\|$ (MHz) | $E$ (MHz) | site orient. | EPR lines | spread (MHz) | $P_{\mathrm{res}}$ | ref |
|---|---|---|---|---|---|---|---|---|---|---|
| $V^-$ (S1) | isolated vacancy, $T_d$ | 3/2 | 1 | 0.0 | 0.0 | 1 | 1 | 0 | 1.000 | 15 |
| P1 ($N_s^0$, $^{14}N$) | substitutional N, $C_{3v}$ | 1/2 | 0 | 0.0 | 0.0 | 4 | 5 | 228 | 0.250 | 21,22 |
| $[V\text{-}V]^0$ (R4/W6) | nearest-neighbour divacancy, $C_{2h}$ | 1 | 2 | 465.0 | −51.5 | 12 | 6 | 929 | 0.187 | 16 |
| R5 | 3V ⟨110⟩ chain | 1 | 3 | 794.6 | −19.6 | 12 | 8 | 813 | 0.125 | 12 |
| $V_3$ analog (B) | 3V chain, low-$T$ $D$ | 1 | 3 | 602.4 | −19.6 | 12 | 8 | 621 | 0.125 | 11,12 |
| O1 | 4V chain | 1 | 4 | 311.1 | −6.2 | 12 | 8 | 317 | 0.125 | 12,13 |
| R6 | 5V chain | 1 | 5 | 181.6 | −2.1 | 12 | 8 | 184 | 0.125 | 12 |
| R10 | 6V chain | 1 | 6 | 113.9 | −0.8 | 12 | 8 | 115 | 0.125 | 12 |
| R11 | 7V chain | 1 | 7 | 76.5 | −0.3 | 12 | 8 | 77 | 0.125 | 12 |
| KUL11 | 8V chain | 1 | 8 | 53.4 | −0.4 | 12 | 7 | 54 | 0.125 | 12 |
| R7a | 9V ring | 1 | 9 | 592.1 | −5.9 | 12 | 8 | 598 | 0.125 | 12 |
| R7 | 11V ring | 1 | 11 | 580.3 | −33.4 | 12 | 12 | 655 | 0.097 | 12 |
| R8 | 13V ring | 1 | 13 | 504.5 | 0.0 | 12 | 4 | 504 | 0.250 | 12 |
| O (isotropic) | isotropic, no fine structure | 1/2 | — | 0.0 | 0.0 | 1 | 1 | 0 | 1.000 | 11 |

**Table 2 | Controlled decomposition of the enhancement.** The same R5 bath at 1 ppm with only the assignment of the fine-structure tensor changed. $T_2$ in µs; at any other concentration $n$ the times scale as $1/n$ and the ratios are unchanged. “Step” is the ratio to the preceding case, with its statistical uncertainty from resampling the configurations.

| case | $T_2$ at 1 ppm (µs) | vs A | step |
|---|---|---|---|
| A. $V^{-}$: $S = 3/2$, no fine structure | 19.1 | **1.00** | — |
| B. $S = 1$, fine structure removed ($D = 0$) | 26.7 | **1.40** | 1.40 ± 0.08 |
| C. $S = 1$, $D$ on, one fixed orientation | 31.8 | **1.67** | 1.19 ± 0.07 |
| D. $S = 1$, $D$ on, 12 equivalent $\langle 110 \rangle$ sites (real) | 73.0 | **3.82** | 2.30 ± 0.20 |

**Table 3 | $NV^{-}$ spin coherence in the defect composition measured after a 1000 °C anneal, read from Fig. 2(b) of ref. 11.** Scenarios 1–7 as described in the Results. P_res is evaluated for the mixture as a whole.

| | scenario | total density (ppm) | $P_{\text{res}}$ | $T_2$ (µs) |
|---|---|---|---|---|
| 1. | measured composition, 1000 °C (chains + O) | 0.96 | 0.601 | 88.4 |
| 2. | all $V^{-}$ at the same total density | 0.96 | 1.000 | 20.0 |
| 3. | all P1 at the same total density | 0.96 | 0.250 | 187 |
| 4. | multi-vacancy chains only, same total density | 0.96 | 0.090 | 97.5 |
| 5. | the measured chains alone (0.22 ppm) | 0.22 | 0.090 | 429 |
| 6. | all $V^{-}$ at the chain density | 0.22 | 1.000 | 87.7 |
| 7. | a single chain species (R5 only), chain density | 0.22 | 0.125 | 336 |

---

# Supplementary Information

The coherence calculations behind every figure and table use dynacce with pair flip-flops of the bath outside each cluster (main text Methods), at $B$ = 300 G along the NV [111] axis and with 50 ppm $^{13}C$, the residual $^{13}C$ level of the $^{12}C$-enriched high-pressure high-temperature crystals used in the author’s own measurements (main text Methods and ref. 24). Concentrations are in ppm of the carbon atom density, 1 ppm = $1.76 \times 10^{17}$ $cm^{-3}$. Coherence times are the 1/e points of the simulated decays; no lifetime is obtained by fitting anywhere in this work.

---

## Supplementary Tables

**Supplementary Table 1 | $T_2$ (Hahn echo, pair flip-flops), µs.**

| centre | $P_{res}$ | 0.01 ppm | 0.1 ppm | 1 ppm | 10 ppm |
|---|---|---|---|---|---|
| $V^-$ (S1) | 1.000 | $1.90 \times 10^3$ | 190 | 19.1 | 1.90 |
| P1 ($N_s^0$, $^{14}N$) | 0.250 | $1.78 \times 10^4$ | $1.79 \times 10^3$ | 179 | 17.9 |
| P1 without hyperfine (control) | 1.000 | $5.19 \times 10^3$ | 519 | 51.9 | 5.19 |
| $[V\text{-}V]^0$ (R4/W6) | 0.187 | $6.71 \times 10^3$ | 672 | 67.1 | 6.72 |
| R5 | 0.125 | $7.67 \times 10^3$ | 768 | 73.0 | 7.35 |
| O1 | 0.125 | $1.09 \times 10^4$ | $1.09 \times 10^3$ | 108 | 10.3 |
| R6 | 0.125 | $1.23 \times 10^4$ | $1.24 \times 10^3$ | 107 | 6.52 |
| R10 | 0.125 | $1.21 \times 10^4$ | $1.21 \times 10^3$ | 116 | 5.16 |
| R11 | 0.125 | $1.21 \times 10^4$ | $1.21 \times 10^3$ | 80.1 | 4.17 |
| KUL11 | 0.125 | $1.16 \times 10^4$ | $1.16 \times 10^3$ | 73.2 | 4.34 |
| R7a | 0.125 | $9.88 \times 10^3$ | 990 | 98.9 | 9.89 |
| R7 | 0.097 | $1.01 \times 10^4$ | $1.01 \times 10^3$ | 101 | 10.1 |
| R8 | 0.250 | $5.67 \times 10^3$ | 567 | 56.7 | 5.67 |

**Supplementary Table 2 | $T_2$ relative to a $V^-$ bath of the same density.**

| centre | 0.01 ppm | 0.1 ppm | 1 ppm | 10 ppm | mean |
|---|---|---|---|---|---|
| P1 ($N_s^0$, $^{14}N$) | 9.38 | 9.44 | 9.39 | 9.43 | **9.41** |
| P1 without hyperfine (control) | 2.73 | 2.73 | 2.72 | 2.73 | **2.73** |
| $[V\text{-}V]^0$ (R4/W6) | 3.53 | 3.53 | 3.52 | 3.53 | **3.53** |
| R5 | 4.04 | 4.04 | 3.82 | 3.87 | **3.94** |
| O1 | 5.73 | 5.74 | 5.65 | 5.42 | **5.63** |
| R6 | 6.49 | 6.50 | 5.61 | 3.43 | **5.51** |
| R10 | 6.35 | 6.36 | 6.08 | 2.71 | **5.38** |
| R11 | 6.36 | 6.37 | 4.19 | 2.19 | **4.78** |
| KUL11 | 6.11 | 6.12 | 3.83 | 2.28 | **4.59** |
| R7a | 5.20 | 5.21 | 5.18 | 5.20 | **5.20** |
| R7 | 5.29 | 5.29 | 5.27 | 5.29 | **5.29** |
| R8 | 2.98 | 2.99 | 2.97 | 2.98 | **2.98** |

**Supplementary Table 3 | $T_2$ from the electron bath alone (μs).** The $^{13}C$ bath changes $T_2$ by at most 0.3 % at every concentration.

| centre | $P_{res}$ | 0.01 ppm | 0.1 ppm | 1 ppm | 10 ppm |
|---|---|---|---|---|---|
| $V^-$ (S1) | 1.000 | $1.90 \times 10^3$ | 190 | 19.1 | 1.90 |
| P1 ($N_s^0$, $^{14}N$) | 0.250 | $1.79 \times 10^4$ | $1.79 \times 10^3$ | 179 | 17.9 |
| P1 without hyperfine (control) | 1.000 | $5.19 \times 10^3$ | 519 | 51.9 | 5.19 |
| $[V\text{-}V]^0$ (R4/W6) | 0.187 | $6.72 \times 10^3$ | 672 | 67.2 | 6.72 |
| R5 | 0.125 | $7.68 \times 10^3$ | 768 | 73.1 | 7.35 |
| O1 | 0.125 | $1.09 \times 10^4$ | $1.09 \times 10^3$ | 108 | 10.3 |
| R6 | 0.125 | $1.24 \times 10^4$ | $1.24 \times 10^3$ | 107 | 6.52 |
| R10 | 0.125 | $1.21 \times 10^4$ | $1.21 \times 10^3$ | 116 | 5.16 |
| R11 | 0.125 | $1.21 \times 10^4$ | $1.21 \times 10^3$ | 80.1 | 4.17 |
| KUL11 | 0.125 | $1.16 \times 10^4$ | $1.16 \times 10^3$ | 73.3 | 4.34 |
| R7a | 0.125 | $9.90 \times 10^3$ | 990 | 99.0 | 9.90 |
| R7 | 0.097 | $1.01 \times 10^4$ | $1.01 \times 10^3$ | 101 | 10.1 |
| R8 | 0.250 | $5.67 \times 10^3$ | 567 | 56.7 | 5.67 |

**Supplementary Table 4 | Annealing series, 800–1000 °C, compositions read from Fig. 2(b) of ref. 11.** For each anneal temperature, the measured composition is compared with an all-$V^-$ bath and an all-multi-vacancy bath of the same total spin density.

| anneal (°C) | case | total density (ppm) | $P_{res}$ | $T_2$ (μs) |
|---|---|---|---|---|
| 800 | measured composition | 4.96 | 0.055 | 48 |
| 800 | same density, all $V^-$ | 4.96 | 1.000 | 3.85 |
| 800 | same density, all multi-vacancy chains | 4.96 | 0.047 | 48.8 |
| 900 | measured composition | 1.61 | 0.294 | 86.6 |
| 900 | same density, all $V^-$ | 1.61 | 1.000 | 11.9 |
| 900 | same density, all multi-vacancy chains | 1.61 | 0.063 | 81.5 |
| 1000 | measured composition | 0.956 | 0.601 | 88.4 |
| 1000 | same density, all $V^-$ | 0.956 | 1.000 | 20 |
| 1000 | same density, all multi-vacancy chains | 0.956 | 0.090 | 97.5 |

**Supplementary Table 5 | Hahn-echo $T_2$ at 1 ppm (electron bath only, μs) with the outer bath frozen (static mean field), flip-flopping in pairs (this work) and jumping independently.**

| centre | static | pair flip-flops | independent jumps | pair / static |
|---|---|---|---|---|
| $V^-$ (S1) | 21.0 | 19.1 | 18.6 | 0.91 |
| P1 ($N_s^0$, $^{14}N$) | 212 | 179 | — | 0.84 |
| P1 without hyperfine (control) | 54.6 | 51.9 | 48.0 | 0.95 |
| $[V\text{-}V]^0$ (R4/W6) | 77.0 | 67.2 | 59.6 | 0.87 |
| R5 | 90.0 | 73.1 | 70.9 | 0.81 |
| O1 | 154 | 108 | 94.9 | 0.70 |
| R6 | 139 | 107 | 88.0 | 0.77 |
| R10 | 143 | 116 | 81.0 | 0.81 |
| R11 | 124 | 80.1 | 72.0 | 0.65 |
| KUL11 | 97.1 | 73.3 | 62.7 | 0.75 |
| R7a | 110 | 99.0 | 83.2 | 0.90 |
| R7 | 136 | 101 | 101 | 0.74 |
| R8 | 64.7 | 56.7 | 52.6 | 0.88 |
| P1, scalar hyperfine model (reference) | 200 | 161 | 142 | 0.81 |
| R5 with $D = 0$ (case B) | 29.2 | 26.7 | 27.5 | 0.91 |
| R5, one orientation (case C) | 33.7 | 31.8 | 29.4 | 0.94 |

**Supplementary Table 6 | $T_2^*$ (FID, signed ensemble average; P1 as an explicit electron–$^{14}N$ unit), μs.**

| centre | $P_{\text{res}}$ | 0.01 ppm | 0.1 ppm | 1 ppm | 10 ppm |
|---|---|---|---|---|---|
| $V^-$ (S1) | 1.000 | 187 | 32.6 | 3.49 | 0.345 |
| P1 ($N_s^0$, $^{14}N$) | 0.250 | 240 | 60.3 | 7.48 | 0.760 |
| P1 without hyperfine (control) | 1.000 | 245 | 58.9 | 6.85 | 0.680 |
| $[V\text{-}V]^0$ (R4/W6) | 0.187 | 218 | 46.6 | 5.28 | 0.540 |
| R5 | 0.125 | 210 | 45.9 | 5.31 | 0.536 |
| O1 | 0.125 | 218 | 47.3 | 5.33 | 0.535 |
| R6 | 0.125 | 217 | 48.1 | 5.41 | 0.547 |
| R10 | 0.125 | 220 | 48.2 | 5.38 | 0.550 |
| R11 | 0.125 | 222 | 48.9 | 5.43 | 0.547 |
| KUL11 | 0.125 | 223 | 49.5 | 5.47 | 0.550 |
| R7a | 0.125 | 211 | 45.0 | 5.12 | 0.526 |
| R7 | 0.097 | 209 | 45.5 | 5.17 | 0.522 |
| R8 | 0.250 | 213 | 45.2 | 5.12 | 0.517 |

**Supplementary Table 7 | $T_2^*$ of the electron bath alone relative to a $V^-$ bath of the same density.**

| centre | 0.01 ppm | 0.1 ppm | 1 ppm | 10 ppm | mean |
|---|---|---|---|---|---|
| P1 ($N_s^0$, $^{14}N$) | 2.22 | 2.18 | 2.16 | 2.20 | **2.19** |
| P1 without hyperfine (control) | 2.03 | 2.00 | 1.99 | 1.97 | **2.00** |
| [V-V]$^0$ (R4/W6) | 1.57 | 1.55 | 1.53 | 1.56 | **1.55** |
| R5 | 1.56 | 1.54 | 1.53 | 1.55 | **1.55** |
| O1 | 1.56 | 1.55 | 1.53 | 1.55 | **1.55** |
| R6 | 1.58 | 1.56 | 1.55 | 1.58 | **1.57** |
| R10 | 1.59 | 1.56 | 1.55 | 1.59 | **1.57** |
| R11 | 1.60 | 1.57 | 1.56 | 1.58 | **1.58** |
| KUL11 | 1.61 | 1.58 | 1.57 | 1.59 | **1.59** |
| R7a | 1.56 | 1.52 | 1.48 | 1.52 | **1.52** |
| R7 | 1.54 | 1.50 | 1.49 | 1.51 | **1.51** |
| R8 | 1.52 | 1.50 | 1.47 | 1.50 | **1.50** |

**Supplementary Table 8 | $T_2^*$ from the electron bath alone (µs).** The $^{13}C$ bath at 50 ppm gives $T_2^*$ = 360 ± 16 µs on its own, so it is irrelevant at 1 and 10 ppm but shortens the total $T_2^*$ at 0.1 and especially 0.01 ppm; Supplementary Table 6 gives the total.

| centre | $P_{\mathrm{res}}$ | 0.01 ppm | 0.1 ppm | 1 ppm | 10 ppm |
|---|---|---|---|---|---|
| $V^-$ (S1) | 1.000 | 345 | 34.9 | 3.52 | 0.346 |
| P1 ($N_s^0$, $^{14}N$) | 0.250 | 765 | 76.2 | 7.62 | 0.762 |
| P1 without hyperfine (control) | 1.000 | 699 | 69.8 | 7.01 | 0.682 |
| [V-V]$^0$ (R4/W6) | 0.187 | 542 | 54.2 | 5.39 | 0.541 |
| R5 | 0.125 | 539 | 53.8 | 5.38 | 0.538 |
| O1 | 0.125 | 539 | 54.0 | 5.39 | 0.536 |
| R6 | 0.125 | 546 | 54.4 | 5.47 | 0.548 |
| R10 | 0.125 | 549 | 54.6 | 5.44 | 0.551 |
| R11 | 0.125 | 551 | 54.9 | 5.48 | 0.548 |
| KUL11 | 0.125 | 554 | 55.2 | 5.51 | 0.551 |
| R7a | 0.125 | 539 | 53.0 | 5.22 | 0.528 |
| R7 | 0.097 | 531 | 52.4 | 5.26 | 0.524 |
| R8 | 0.250 | 522 | 52.4 | 5.18 | 0.519 |

**Supplementary Table 9 | $T_2^*$ of the measured compositions and the baths compared with them (µs).** Annealing series as in Supplementary Table 4, scenarios as in Table 3 of the main text.

| bath | total density (ppm) | $T_2^*$ (µs) |
|---|---|---|
| 800 °C, measured composition | 4.96 | 1.04 |
| 800 °C, same density, all $V^-$ | 4.96 | 0.695 |
| 800 °C, same density, all multi-vacancy chains | 4.96 | 1.01 |
| 900 °C, measured composition | 1.61 | 3.36 |
| 900 °C, same density, all $V^-$ | 1.61 | 2.12 |
| 900 °C, same density, all multi-vacancy chains | 1.61 | 2.77 |
| 1000 °C, measured composition | 0.956 | 6.72 |
| 1000 °C, same density, all $V^-$ | 0.956 | 3.54 |
| 1000 °C, same density, all multi-vacancy chains | 0.956 | 4.56 |
| scenario 3, all P1 | 0.956 | 7.84 |
| scenario 5, the measured chains alone | 0.218 | 19.4 |
| scenario 6, all $V^-$ at the chain density | 0.218 | 15.3 |
| scenario 7, R5 only at the chain density | 0.218 | 22.8 |

## Supplementary Note 1. Spin-Hamiltonian parameters

Table 1 of the main text lists the parameters actually used. Their provenance and the unit conversions applied to them are as follows.

**Isolated negative vacancy $V^-$ (S1 centre).** $S = 3/2$, $g = 2.0027$ isotropic, T_d symmetry. Isoya et al.[15] state explicitly that fine-structure splittings vanish for $S = 3/2$ in T_d, so $\mathbf{D} = 0$ exactly; the effective spin was determined from the $^{13}C$ ENDOR frequencies of the next-nearest-neighbour shell. This is the reference “no frequency structure” bath of this work.

**Neutral divacancy $[V\text{–}V]^0$ (R4/W6).** $S = 1$, $C_{2h}$ at 33 K, principal values of the traceless **D** matrix $D_1 = +103$, $D_2 = +206$, $D_3 = -310$ MHz, with $D_3$ at 54.2° from [001] in the (110) plane, i.e. essentially along ⟨111⟩.[16] Taken directly in MHz.

**⟨110⟩ vacancy chains and closed clusters (R5, O1, R6, R10, R11, KUL11, R7a, R7, R8).** $S = 1$. Iakoubovskii and Stesmans[12] report the principal values $D_1$, $D_2$, $D_3$ of the traceless **D** matrix in field units (mT). Every centre was checked for $D_1 + D_2 + D_3 = 0$, then converted with $\nu[\mathrm{MHz}] = g \times 13.9962 \times D[\mathrm{mT}]$. The principal axis of **D** is ⟨110⟩. For R7a the two minor axes lie along ⟨1-10⟩ and ⟨001⟩; for R7 they are tilted by 15°. Both cases are implemented.

**P1 ($N_s^0$).** $S = 1/2$, $I = 1$ ($^{14}N$), $A\| = 114.03$ MHz, $A\perp = 81.31$ MHz, quadrupole $P = -5.01$ MHz, Jahn–Teller axis along one of the four ⟨111⟩ directions.[21],[22]

**Measured composition.** Yamamoto et al.[11] resolve seven $S = 1$ multi-vacancy chain signals (A–G) plus an isotropic broad signal O. A, D, E, F and G were identified with $V_3$ (R5), $V_4$ (O1), $V_5$ (R6), $V_6$ (R10) and $V_7$ (R11) respectively, and the published parameters of those centres are used here. B and C are further $V_3$ variants: B is stated to have almost the same fine-structure splitting as R5 at low temperature (≤ 77 K), so it is given the low-temperature $D$ of R5 (21.5 mT, from Fig. 4 of ref. 12, against 28.35 mT at 300 K), while C has no reported splitting and is given the room-temperature R5 parameters. C is treated as identical to A rather than as a distinct species. That choice is conservative: it *underestimates* the number of distinct transition frequencies in the mixture, and therefore underestimates the mixture effect discussed in the Results. The isotropic signal O is modelled as an $S = 1/2$ centre with isotropic $g = 2.0028$ and no fine structure, the natural description of a carbon dangling bond; its spin quantum number is not stated in ref. 11 and this is an assumption. Concentrations are read from Fig. 2(b) of ref. 11, which gives them in $cm^{-3}$, with an accuracy of roughly 30 %; the values used, converted to ppm, are

| anneal | chains (ppm) | isotropic O (ppm) | total (ppm) | chain species |
|---|---|---|---|---|
| 800 °C | 4.26 | 0.71 | 4.96 | 6 |
| 900 °C | 0.76 | 0.85 | 1.61 | 6 |
| 1000 °C | 0.22 | 0.74 | 0.96 | 3 |

For each temperature the sum over A–G and the total were checked against the corresponding curves in the same figure. The per-species concentrations used as input are

| anneal | R5 ($V_3$; A and C) | $V_3$ analog (B) | O1 ($V_4$) | R6 ($V_5$) | R10 ($V_6$) | R11 ($V_7$) | isotropic O |
|---|---|---|---|---|---|---|---|
| 800 °C | 2.02 | 0.074 | 1.59 | 0.43 | 0.12 | 0.019 | 0.71 |
| 900 °C | 0.52 | 0.016 | 0.11 | 0.068 | 0.040 | 0.0045 | 0.85 |
| 1000 °C | 0.18 | 0.0074 | 0.028 | — | — | — | 0.74 |

all in ppm. Signals A and C are treated as one species with the room-temperature R5 parameters, as explained above. The echo for each composition is computed at a total density of 1 ppm with these species fractions and transferred to the measured total density by the $1/n$ scaling (main text Methods); the free induction decay is computed directly at the measured total density.

Only the 800, 900 and 1000 °C points are modelled, although ref. 11 reports spectra up to 1200 °C. At 1100 °C and above the chain signals A–G have annealed out and two further signals dominate, denoted H and R12. H is assigned only tentatively — ref. 11 describes it as a multi-vacancy cluster without the $C_{2v}$ symmetry of the chain structures — and no fine-structure tensor has been reported for it. R12 ($C_{3v}$, $S$ = 1) has published spin-Hamiltonian parameters, but ref. 11 notes that it overlaps the isotropic signal O, so the two concentrations cannot be separated in Fig. 2(b) of that work. A centre whose fine-structure tensor is unknown cannot be assigned one without prejudging P_res, which is the quantity under study; and a concentration that cannot be separated from another species cannot be used as an input. The high-temperature points are therefore left out rather than guessed at.

---

## Supplementary Note 2. Validity of the high-field reduction for P1

A P1 centre is a two-spin object (electron + $^{14}N$). A spin-1/2 electron with a scalar detuning $\delta = m_I A_{zz}(\theta)$ is a high-field approximation to that object, and it fails at low field. Its validity at the field used here was therefore checked by exact diagonalisation of

$H = g\mu_B \mathbf{B}\cdot\mathbf{S} - \gamma_N \mathbf{B}\cdot\mathbf{I} + \mathbf{S}\cdot\mathbf{A}\cdot\mathbf{I} + \mathbf{I}\cdot\mathbf{P}\cdot\mathbf{I}$,

with **A** and **P** axial about the centre's own Jahn–Teller axis. The expectation value of the electron spin along the field, $\langle S_B \rangle$, taken over all six eigenstates and all four Jahn–Teller orientations, is

| $B$ (G) | $\langle S_B \rangle$ range | scalar reduction |
|---|---|---|
| 25 | 0.035 – 0.500 | fails (electron–nuclear hybridised) |
| 100 | 0.434 – 0.500 | marginal |
| **300 (this work)** | **0.493 – 0.500** | **valid** |
| 500 | 0.497 – 0.500 | valid |
| 1000 | 0.499 – 0.500 | valid |

At 25 G the electron and its nucleus hybridise, and the scalar reduction is not usable. At 300 G the exact transition frequencies differ from the scalar ones by at most 11 MHz. That second-order shift is common to every member of a degeneracy class, so the five-way structure of the P1 spectrum — and hence the resonant fraction P_res = 1/4 — is unchanged. The smallest mismatch between classes, 28 MHz, stays about three orders of magnitude above the dipolar coupling.

---

## Supplementary Note 3. The resonant flip-flop fraction and its threshold

P_res is defined as the weighted probability that two bath spins, drawn independently with their site orientations and nuclear states (and, for a multi-species bath, with their species drawn in proportion to concentration), present EPR transition frequencies agreeing to within a threshold. The threshold must sit above the bath–bath dipolar coupling and below the smallest structural splitting the bath presents. Over the concentrations studied the electron–electron dipolar coupling at the mean spacing is

| concentration (ppm) | mean spacing (nm) | dipolar coupling (MHz) |
|---|---|---|
| 0.01 | 83 | $9 \times 10^{-5}$ |
| 0.1 | 38 | $9 \times 10^{-4}$ |
| 1 | 18 | $9 \times 10^{-3}$ |
| 10 | 8.3 | $9 \times 10^{-2}$ |

and the smallest structural splittings in Table 1 are 47 kHz for KUL11 and 137 kHz for R11 (the nearly degenerate site orientations). A threshold of 10 kHz is used throughout; every threshold between 1 and 40 kHz returns identical P_res for every centre, whereas KUL11 doubles to 0.250 from 50 kHz and R11 rises to 0.312 from 150 kHz. No single threshold meets both conditions at 10 ppm, where the dipolar coupling reaches 0.09 MHz, and the long chains lose their enhancement at high concentration (main text Results and Fig. 3b; Supplementary Note 4). A 1 MHz threshold would instead assign R6, R10, R11 and KUL11 P_res of 0.19, 0.44, 0.50 and 0.50 rather than 0.125. The threshold is therefore stated explicitly.

---

## Supplementary Note 4. Convergence and protocol

*Cutoffs.* For the Hahn echo r_dipole = 4 $n$^(−1/3) and r_bath = 5.5 $n$^(−1/3) (below). Because both scale as $n$^(−1/3), each concentration in the series has the same number of spins and the same number of pairs, so that the concentration dependence reported is physical and not a convergence trend.

*Order.* The Hahn echo is computed at CCE order 2. Order 3 could not serve as a convergence test. CCE beyond order 2 is known to turn unphysical in dilute dipolar electron baths, because a lower-order cluster factor passing near zero enters a higher-order term as a denominator, while order 2 stays in the physical range (main-text refs 25, 26). In two configurations each of $V^-$ and P1 (fully mixed, r_dipole = 1.2 $n$^(−1/3), $1 \times 10^{17}$ $cm^{-3}$), CCE3 departs from CCE2 by 10 % at 0.5–1.2 $T_2$ and then grows to $|L| = 2 \times 10^{29}$ to $9 \times 10^{203}$, which no coherence can do. These runs carry no mean field from the spins outside each cluster. CCE3 lies above CCE2 at $T_2$ in all four, the sign reported for an expansion averaged over bath states with that field included: odd orders lengthen the decay and even orders shorten it, by corrections of the same order of magnitude (main-text ref. 26). The gap between CCE3 and CCE2 alone therefore need not measure the error of CCE2. Convergence in the order is nevertheless not established in this work, and the support for CCE2 is external: the comparison with the measured P1 $T_2$, and the published tests described in Methods (main-text refs 25, 26). Those tests do not cover a bath of $S = 3/2$, so the $T_2$ of $V^-$, and the ratios taken to it, may carry a truncation error that this work does not quantify.

*Ensemble size.* 12 spatial configurations, each with four sampled bath states.

*Echo envelope.* Conventional CCE evaluates a ratio of cluster to subcluster contributions independently at each time point, so isolated points can exceed unity when a subcluster contribution passes through zero. A Hahn-echo envelope in this bath has no revival mechanism, so the running minimum of each echo decay is taken before the 1/e point is read off. The envelope changes the extracted 1/e time by less than 1 % for every curve in this work, so no result depends on it.

*Bath state and cluster cutoff.* A fully mixed bath has no mean field from the spins outside each cluster, and its $T_2$ is too short. The production runs sample each bath spin in one eigenstate of its single-spin Hamiltonian (four samples per configuration), so that dynacce adds the mean field of the outer spins, and use r_dipole = 4 $n$^(−1/3) and r_bath = 5.5 $n$^(−1/3). $T_2$ at $1 \times 10^{17}$ $cm^{-3}$ with the outer bath frozen (electron bath only, μs; eight configurations; uncertainties from resampling the configurations):

| centre | fully mixed, r_dipole = 1.2 | sampled, 3 | sampled, 4 | sampled, 5 |
|---|---|---|---|---|
| $V^-$ | 26.1 | 43.0 ± 5.0 | 42.7 ± 5.0 | 44.7 ± 6.5 |
| R5 | 110 | 171 ± 26 | 162 ± 23 | 155 ± 22 |
| $[V\text{–}V]^0$ | 96.5 | 135 ± 34 | 132 ± 30 | 185 ± 15 |
| P1 (scalar model) | 174 | 386 ± 61 | 376 ± 46 | 384 ± 49 |

r_dipole in units of $n$^(−1/3). Going from 4 to 3 mean spacings changes $T_2$ by factors of 1.01–1.06 (each ± 0.07 or less). Going from 4 to 5 changes it by 1.05 ± 0.14 ($V^-$), 0.96 ± 0.16 (R5), 1.40 ± 0.28 ($[V\text{–}V]^0$) and 1.02 ± 0.19 (P1). These lie within the uncertainty, but eight configurations cannot exclude a change of 20–30 % at 5 mean spacings, and the $[V\text{–}V]^0$ value in particular is not settled. The fully mixed bath gives a $T_2$ shorter than the 4-spacing value by 1.64 ± 0.23 ($V^-$), 1.47 ± 0.32 (R5), 1.37 ± 0.39 ($[V\text{–}V]^0$) and 2.16 ± 0.49 (P1). The same runs repeated with PyCCE (main-text ref. 23) agree with dynacce to better than one part in $10^8$. With a cutoff near one spacing it freezes (not shown), because the only clusters left are close pairs that the mean field detunes. The same behaviour was found for a static P1 bath by Park et al. (main-text ref. 20).

*Pair flip-flops of the outer bath.* Each outer spin flip-flops with partners within 4 mean spacings, and each of the four sampled states per configuration starts one stochastic trajectory. Changing the partner range to 3 or 5 mean spacings, or the number of trajectories to 8, changes $T_2$ at 1 ppm as follows (ratios to the production value):

| bath | production $T_2$ (μs) | 3 spacings | 5 spacings | 8 trajectories |
|---|---|---|---|---|
| $V^-$ | 19.1 | 0.99 ± 0.04 | 1.00 ± 0.05 | 0.96 ± 0.05 |
| P1 (scalar model) | 161 | 1.00 ± 0.06 | 0.89 ± 0.04 | 0.95 ± 0.08 |
| R5 | 73.1 | 1.07 ± 0.06 | 1.08 ± 0.07 | 1.04 ± 0.06 |
| 1000 °C, measured composition | 84.5 | 0.93 ± 0.05 | 0.91 ± 0.07 | 0.90 ± 0.03 |
| 1000 °C, chains only | 93.3 | 1.15 ± 0.07 | 1.03 ± 0.08 | 1.05 ± 0.06 |

There is no monotonic trend with the range. A dependence on these settings of about the size of the statistical uncertainty remains, and it is not included in the uncertainties quoted elsewhere. With the outer bath frozen, 8 trajectories give 1.01 ± 0.03, 1.01 ± 0.07, 1.03 ± 0.06, 0.99 ± 0.08 and 0.91 ± 0.07 in the same order. The explicit P1 unit was not part of this check.

*Concentration scaling.* The echo is computed at 1 ppm and transferred to the other concentrations as $L_n(t) = L(tn/1\text{ ppm})$, which is exact when only the dipolar coupling changes with $n$. Direct calculations test it ($T_2$ direct / $T_2$ scaled from 1 ppm):

| centre | static, 0.1 ppm | static, 10 ppm | pair, 0.1 ppm | pair, 10 ppm |
|---|---|---|---|---|
| $V^-$ | 1.001 ± 0.002 | 1.002 ± 0.002 | 0.995 ± 0.018 | 0.996 ± 0.018 |
| R5 | 0.984 ± 0.019 | 0.982 ± 0.019 | 1.05 ± 0.04 | 1.01 ± 0.07 |
| R5, $D = 0$ | 1.000 ± 0.001 | 1.001 ± 0.001 | 1.02 ± 0.03 | 1.02 ± 0.03 |
| O1 | 0.994 ± 0.021 | 0.925 ± 0.038 | 1.01 ± 0.09 | 0.96 ± 0.07 |
| R6 | 1.03 ± 0.06 | 0.74 ± 0.09 | 1.15 ± 0.16 | 0.61 ± 0.06 |
| R10 | 1.08 ± 0.04 | 0.46 ± 0.07 | 1.04 ± 0.11 | 0.44 ± 0.04 |
| R11 | 1.21 ± 0.12 | 0.36 ± 0.03 | 1.51 ± 0.17 | 0.52 ± 0.05 |
| KUL11 | 1.48 ± 0.20 | 0.52 ± 0.06 | 1.59 ± 0.09 | 0.59 ± 0.03 |

The direct values are used in all tables and figures for these centres, and 0.01 ppm is obtained from 0.1 ppm. The other centres, whose near-resonant channels are absent or negligible, and the measured compositions are transferred from 1 ppm without a direct test. The failure appears with a frozen outer bath as well, so the flip-flops of the outer bath are not its main cause. R5 and its $D = 0$ control show that a large $D$ alone does not break the scaling.

*Single-orientation control.* Assigning every defect the tensor of one site restores the scaling ($T_2 \cdot n$ at 10 ppm / $T_2 \cdot n$ at 0.1 ppm):

| centre | twelve orientations, static | twelve orientations, pair | one orientation, static | one orientation, pair |
|---|---|---|---|---|
| KUL11 | 0.354 ± 0.045 | 0.374 ± 0.023 | 0.995 ± 0.009 | 0.999 ± 0.003 |
| R11 | 0.299 ± 0.042 | 0.345 ± 0.035 | 0.992 ± 0.015 | 0.998 ± 0.004 |

This shows that different orientations are necessary. It does not isolate the near-resonant channels, because fixing the orientation also changes every coupling between differently oriented sites and the distribution of couplings to the $NV^-$. The dependence on which site is chosen was not examined.

*Field control.* $T_2 \cdot n$ at 600 G / $T_2 \cdot n$ at 300 G:

| comparison (600 G / 300 G) | KUL11, static | R11, static | KUL11, pair | R11, pair |
|---|---|---|---|---|
| 1 ppm / 1 ppm | 0.73 ± 0.08 | 0.72 ± 0.09 | 0.83 ± 0.05 | 0.78 ± 0.07 |
| 0.25 ppm / 1 ppm (same $nB^2$) | 1.08 ± 0.04 | 1.10 ± 0.07 | 1.08 ± 0.06 | 1.36 ± 0.10 |
| 2.5 ppm / 10 ppm (same $nB^2$) | 1.41 ± 0.13 | 1.41 ± 0.09 | 1.27 ± 0.08 | 1.19 ± 0.05 |
| one orientation, 1 ppm / 10 ppm | 1.01 ± 0.02 | 1.00 ± 0.02 | 1.13 ± 0.07 | 1.05 ± 0.05 |

For a ⟨110⟩ chain with the field along [111], the $\cos^2\theta$ of the $D$ axis is 0 or 2/3, so the first-order transitions of the two groups of sites coincide. Single-defect diagonalisation of the fine-structure and Zeeman terms then gives two groups of flip-flop detunings. One, a third of the weight, follows about 0.225 $D^3/(\gamma B)^2$, a coefficient found numerically (0.223–0.235 for $D$ = 53–311 MHz at 300 and 600 G) and not derived; it moves from 47 to 12 kHz for KUL11 and from 140 to 35 kHz for R11 between 300 and 600 G. The other, two thirds of the weight, stays near $|E|$ at either field (about 420 kHz for KUL11 and 280 kHz for R11). At a fixed concentration the higher field shortens $T_2 \cdot n$, as closer sites should, and a frozen single-orientation bath does not respond. At equal $nB^2$ the ratios exceed 1, so the first group alone does not decide the outcome. That the second group contributes is an inference from those ratios; it was not tested on its own. The pair value 1.13 ± 0.07 for one orientation of KUL11 is unexplained.

*Limits of this reading.* It is not a quantitative model. The field test covers KUL11 and R11 only, and across the five chains the evidence is a rank order: the near-resonant weight within 1 MHz, relative to the exactly resonant weight, is 0.004, 0.5, 2.5, 3.0 and 3.0 for O1, R6, R10, R11 and KUL11. The calculation uses isotropic $g$ values; a $g$ tensor changes the size of these detunings. It omits the inhomogeneous linewidth of real samples ($^{13}C$ hyperfine, strain), in which detunings of a few hundred kHz or less may be buried, and the field is exactly along [111], so the effect of a tilt is unknown. $T_2$ is read at the 1/e point and the shapes of these curves were not examined in detail.

---

## Supplementary Note 5. Validation against the P1 literature

| $T_2$ slope (µs·ppm) | this work | experiment (ref. 8) |
|---|---|---|
| pair flip-flops, explicit electron–$^{14}N$ unit (this work) | 179 ± 22 | 160 ± 12 |
| outer bath frozen, explicit unit | 212 ± 39 | — |
| pair flip-flops, scalar model | 161 ± 13 | — |
| outer bath frozen, scalar model | 200 ± 23 | — |
| independent jumps, scalar model | 142 ± 20 | — |
| hyperfine off | 52 ± 4 | — |

The pair flip-flop calculation lies 12 % above the measured slope, 0.8 combined standard deviations, with no adjustable parameter. Its ensemble curve falls through 1/e without rising again: the first and last crossings both give 179 µs·ppm, and a stretched-exponential fit, used only as a check, gives 178 µs·ppm (exponent 1.10). Freezing the outer bath gives 33 % above the measurement, 1.3 combined standard deviations, in line with independent converged static calculations that give 236–286 µs·ppm (main-text ref. 20). That value is less well defined, because the frozen curve dips and rises near 1/e, with a first crossing at 212 and a last at 244 µs·ppm. Pair flip-flops shorten $T_2$ by 1.18 ± 0.15. The explicit unit and the scalar model, on the same positions, differ by 1.11 ± 0.13 with pair flip-flops and 1.06 ± 0.23 frozen. Removing the on-site $^{14}$N hyperfine alone costs a factor 3.5 ± 0.3 in $T_2$ — the same mechanism this paper studies, acting within the P1 bath itself. All uncertainties come from resampling the 12 configurations. $T_2^*$ is compared with the measurement in Supplementary Note 7.

---

## Supplementary Note 6. Frozen, pair-flip-flop and independently jumping outer bath

Supplementary Table 5 compares, at 1 ppm, the three treatments of the spins outside each cluster available in dynacce: frozen in their sampled states (static mean field), flip-flopping in pairs (the treatment used throughout this work and validated on P1 in the main text), and jumping independently of a partner. Pair flip-flops shorten $T_2$ by factors of 0.65–0.95. The shortening is smallest where the bath has no frequency structure to protect ($V^-$, 0.91; the hyperfine-free P1 control, 0.95; case B of the decomposition, 0.91), and it reaches 0.65–0.90 for the multi-vacancy centres. With the outer bath frozen, the enhancements over $V^-$ would be 3.1–7.3 instead of 3.0–6.1. The decomposition hardly depends on the treatment: frozen and pair flip-flops give 1.15 ± 0.09 and 1.19 ± 0.07 for the single fixed orientation of case C, and 2.67 ± 0.23 and 2.30 ± 0.20 for the twelve real orientations. Replacing the isotropic signal with chains after a 1000 °C anneal gives 1.30 ± 0.29 with a frozen outer bath and 1.10 ± 0.13 with pair flip-flops, in neither case distinguishable from no change. The 800 °C composition exceeds a P1 bath of equal spin density under both, by 1.49 ± 0.34 and 1.33 ± 0.17. The qualitative conclusion — that a multi-vacancy bath is far more benign than a bath of isolated negative vacancies at the same spin density, and that the site orientations are mainly responsible — holds under all the treatments for which it was tested.

---

## Supplementary Note 7. Free induction decay

The main text concerns $T_2$. $T_2^*$ of the same baths is collected here (Supplementary Tables 6–9). It is computed at CCE order 1, where the expansion is exact for a static bath, in a fully mixed bath with $\langle N \rangle \approx 300$ spins inside r_bath (3440, 1597, 741 and 344 Å at 0.01, 0.1, 1 and 10 ppm) and 192 configurations; it self-averages much more slowly than the echo, because a given $NV^-$ is dominated by its nearest bath spin and that distance is broadly distributed. The complex coherence is averaged over the configurations before its magnitude is taken, as in a measured ensemble signal; averaging the magnitude configuration by configuration would make the decay too slow. P1 is treated here as an explicit electron–$^{14}$N unit, as in the main text; the scalar model gives 7.5 ± 0.7 µs·ppm.

*Comparison between baths.* $T_2^*$ orders the baths by spin quantum number, whatever $D$, $E$ or P_res, and that of the electron bath follows $1/n$ to within a few per cent for every bath. At 1 ppm the ratio to $V^-$ is 2.16 ± 0.24 for P1, 1.99 ± 0.12 for the hyperfine-free control and 1.47–1.57 (each ± about 0.15) for every $S = 1$ species. In a dilute dipolar bath the ensemble decay rate is proportional to the mean magnitude $\langle |m| \rangle$ of the bath spin projection, which is 1 for $S = 3/2$, 2/3 for $S = 1$ and 1/2 for $S = 1/2$, and so predicts ratios of 1.5 and 2, in agreement with all three. The differences among the $S = 1$ centres lie within the statistical uncertainty, and so does the effect of the $^{14}$N hyperfine on $T_2^*$ (a ratio of 1.09 ± 0.15 to the hyperfine-free control).

*Comparison with measurement.* The electron bath gives a P1 slope of 7.6 ± 0.7 µs·ppm, 21 % below the measured 9.6 ± 0.9 µs·ppm (ref. 8), a difference of 1.8 combined standard deviations with no adjustable parameter. Other dephasing sources in a real crystal, such as strain and further paramagnetic spins, would shorten the measured value further, so they cannot account for the gap, whose cause is not established here.

*The ratio $T_2/T_2^*$.* $T_2^*$ carries no flip-flop dynamics, so $T_2/T_2^*$ isolates flip-flop blocking. At 1 ppm it is 5.5 for $V^-$, 7.6 for the hyperfine-free control and 11.1–21.6 for the multi-vacancy centres. For P1 it is 24, against the ≈ 17 that follows from the measured slopes (160 and 9.6 µs·ppm, ref. 8).

*The $^{13}$C bath.* On its own the $^{13}$C bath at 50 ppm gives $T_2^* = 360 \pm 16$ µs, from 768 configurations; a bath radius of 250 Å instead of 350 Å gives 345 ± 16 µs, the same within the statistical uncertainty. It is irrelevant at 1 and 10 ppm and shortens the total at 0.1 ppm and below, for example the $V^-$ value at 0.01 ppm from 345 to 187 µs.

*Ensemble size.* Convergence of the P1 $T_2^*$ at 1 ppm with the number of configurations:

| configurations | explicit unit | scalar model | scalar model, mean of $|L|$ |
|---|---|---|---|
| 16 | 7.9 | 7.8 | 9.9 |
| 48 | 7.4 | 7.4 | 11.5 |
| 96 | 7.9 | 7.6 | 11.7 |
| 192 | 7.6 | 7.5 | 11.5 |

$T_2^*$ in μs, electron bath only. The last column averages the magnitude configuration by configuration and is shown only to give the size of that bias.

---

## Supplementary Note 8. Which baths need exact pair propagators

A two-spin cluster in the time-dependent mean field can be propagated exactly, step by step, or in a secular approximation. The secular approximation works in the eigenbasis of the cluster Hamiltonian, keeps the mean-field couplings only among nearly degenerate levels, and carries the couplings between other levels as a second-order energy shift. What it drops is a shift of order $b^3/\Delta^2$, with $b$ the mean field and $\Delta$ the level spacing, and that shift accumulates as a phase. The mean field acts through S_z, so the error vanishes when S_z commutes with the bath spin Hamiltonian, and otherwise grows with the part of S_z that connects different eigenstates. The weight $w$ below measures that part: $w = 1 - \Sigma_k \langle k|S_z|k\rangle^2/\mathrm{Tr}\, S_z^2$, over the eigenstates of one $S = 1$ bath spin at 300 G, averaged over its site orientations.

All values in this work use exact propagators. The table shows what the secular approximation would have given for $T_2$ at 1 ppm (electron bath only). “Ratio” is exact / secular for the ensemble $T_2$, with its statistical uncertainty from resampling the configurations; “configurations” gives the range of that ratio over single configurations.

| bath | $\|D\|$ (MHz) | $w$ (%) | secular (μs) | exact (μs) | ratio | configurations |
|---|---|---|---|---|---|---|
| $V^-$ | 0 | 0 | 19.1 | 19.1 | 1.00 | 1.00 |
| P1 (scalar) | 0 | 0 | 161 | 161 | 1.00 | 1.00 |
| P1, no hyperfine | 0 | 0 | 51.9 | 51.9 | 1.00 | 1.00 |
| R5, case B | 0 | 0 | 26.7 | 26.7 | 1.00 | 1.00 |
| KUL11 | 53 | 0.10 | 73.2 | 73.3 | 1.00 ± 0.00 | 0.99–1.01 |
| R11 | 76 | 0.21 | 80.1 | 80.1 | 1.00 ± 0.01 | 0.99–1.02 |
| R10 | 114 | 0.46 | 116 | 116 | 1.00 ± 0.01 | 0.66–1.08 |
| R6 | 182 | 1.2 | 105 | 107 | 1.02 ± 0.06 | 0.94–1.10 |
| O1 | 311 | 3.5 | 95.9 | 108 | 1.12 ± 0.10 | 0.59–2.16 |
| $[V–V]^0$ | 465 | 6.5 | 62.4 | 67.2 | 1.08 ± 0.06 | 0.70–1.13 |
| R8 | 505 | 9.2 | 50.3 | 56.7 | 1.13 ± 0.14 | 0.97–4.74 |
| R7 | 580 | 12 | 86.6 | 101 | 1.16 ± 0.19 | 0.86–9.63 |
| R7a | 592 | 12 | 73.8 | 99.0 | 1.34 ± 0.19 | 0.96–4.65 |
| R5 | 795 | 21 | 64.1 | 73.1 | 1.14 ± 0.15 | 0.97–2.10 |
| R5, case C | 795 | 21 | 24.6 | 31.8 | 1.29 ± 0.19 | 0.96–2.63 |
| 800 °C chains | — | — | 217 | 242 | 1.11 ± 0.13 | 0.91–6.40 |
| 900 °C chains | — | — | 120 | 132 | 1.10 ± 0.06 | 0.88–1.88 |
| 1000 °C chains | — | — | 83.7 | 93.3 | 1.12 ± 0.08 | 0.99–1.95 |
| 800 °C measured | — | — | 231 | 238 | 1.03 ± 0.03 | 0.99–1.18 |
| 900 °C measured | — | — | 140 | 140 | 1.00 ± 0.01 | 0.97–1.00 |
| 1000 °C measured | — | — | 84.5 | 84.5 | 1.00 ± 0.00 | 1.00 |

The baths fall into three groups.

*S_z commutes with the bath spin Hamiltonian ($V^-$, the scalar P1 model, the hyperfine-free control, case B).* The two propagators give the same result to within $7 \times 10^{-5}$ in the coherence. The approximation is sufficient here, for the reason given above and not by chance. P1 as an explicit electron–$^{14}N$ unit does not commute exactly, because the hyperfine interaction mixes the electron and the nucleus slightly ($w$ = 1.6 % over its six eigenstates). A direct test of that unit at 300 G, on 12 configurations of another position set, gives exact / secular = 1.001 ± 0.001, with at most 0.015 change in the coherence, so the approximation suffices for it too.

*$S$ = 1 centres with small splittings (KUL11, R11, R10, R6; $|D| \leq$ 182 MHz, $w \leq$ 1.2 %).* The ensemble $T_2$ agrees to within 2 %, so the approximation is sufficient for the ensemble value at 1 ppm. Single configurations can still differ, by up to a third for R10.

*$S$ = 1 centres with large splittings (O1 and above; $|D| \geq$ 311 MHz, $w \geq$ 3.5 %).* These need exact propagators. The secular approximation shortens the ensemble $T_2$ by factors of 1.08–1.34. Each of those ratios lies only one or two standard deviations above 1, but all lie on the same side, and single configurations change by up to a factor 9.6 (R7). The size of the effect follows $w$ as a threshold and not in detail: R7a changes more than R5, although R5 has the larger $w$. The errors would also have changed the ranking of the centres, because they differ from centre to centre: R7a would have appeared at 3.87 times $V^-$ instead of 5.18.

Mixtures follow their components. The chains alone change by about 10 %, whereas the measured composition at 1000 °C, in which 77 % of the spins are the isotropic $S$ = 1/2 signal, does not change.

Two extrapolations follow from the form of the error, and neither was tested on the multi-vacancy baths. Since $w$ falls as the Zeeman energy grows relative to $D$, a higher field should enlarge the range of baths for which the approximation suffices; the P1 unit shows that trend, with exact / secular = 1.022 ± 0.008 at 25 G against 1.001 ± 0.001 at 300 G. Since $b$ grows with the concentration $n$ while $T_2$ falls as $1/n$, the accumulated phase at a time of order $T_2$ should grow with concentration, so that the approximation should be safer below 1 ppm than above it.